\documentclass[amsmath,amssymb,11pt]{article}
\usepackage{jheppub2}
\pdfoutput=1
\usepackage{graphicx,epsfig}
\usepackage{float}
\usepackage{subfig}
\usepackage{subfloat}
\usepackage[normalem]{ulem}

\usepackage[utf8]{inputenc}
\usepackage{hyperref}
\usepackage{caption}
\usepackage{color}
\usepackage{overpic}
\usepackage[dvipsnames]{xcolor}

\newcommand*{\affmark}[1][*]{\textsuperscript{#1}}

\definecolor{pink1}{rgb}{0.858, 0.188, 0.478}

\newcommand{\beq}{\begin{equation}}

\newcommand{\eeq}{\end{equation}}

\title{Holographic CFT Phase Transitions  and  Criticality for Charged AdS Black Holes}

\author{Wan Cong,\affmark[1]}
\emailAdd{wan.cong@univie.ac.at}
\author{David Kubiz\v{n}\'{a}k,\affmark[3,2,4]}
\emailAdd{dkubiznak@perimeterinstitute.ca}
\author{Robert B. Mann\affmark[2,3]}
\emailAdd{rbmann@uwaterloo.ca}
\author{and Manus R. Visser\affmark[5]}
\emailAdd{manus.visser@unige.ch}

\affiliation{\affmark[1]University of Vienna, Faculty of Physics,\\
Vienna, Austria\\ 
\affmark[2]Department of Physics and Astronomy, University of Waterloo,\\
 Waterloo, Ontario, N2L 3G1, Canada\\ 
\affmark[3]Perimeter Institute for Theoretical Physics, \\Waterloo, Ontario, N2L 2Y5, Canada\\ 
\affmark[4]Institute of Theoretical Physics, Faculty of Mathematics and Physics, Charles University, \\ 
Prague, V Hole\v{s}ovi\v{c}k\'{a}ch 2, 180 00 Prague 8, Czech Republic\\ 
\affmark[5]Department of Theoretical Physics, University of Geneva,\\
 24 quai Ernest-Ansermet, 1211 Gen\`{e}ve~4, Switzerland\\}

\abstract{We study the
 holographic dual of the 
extended thermodynamics of spherically symmetric, charged AdS black holes in the context of the AdS/CFT correspondence. 
The gravitational   thermodynamics of AdS black holes can be extended by allowing for variations of the cosmological constant and  Newton's constant.
In the  dual CFT this corresponds to   including the central charge $C$ and its chemical potential $\mu$ as a new pair of conjugate thermodynamic  variables, in  addition to the  standard   pairs:   temperature vs. entropy $(T,S)$, electric potential vs.   charge $(\tilde \Phi, \tilde Q)$ and   field theory pressure vs. volume $(p,{\cal V})$. 
For the  (grand) canonical ensembles at fixed $(\tilde Q, {\cal V}, C), (\tilde \Phi, {\cal V},C)$ and $(\tilde Q, {\cal V}, \mu)$, based on the holographic dictionary,  we argue the CFT description of charged AdS black holes contains  either critical phenomena or interesting phase behaviour. 
In the fixed $(\tilde Q, \mathcal V, \mu)$ we find a new zeroth-order phase transition between  a high- and low-entropy phase at some $\mu$-dependent temperature. Finally, we point out there is no critical behaviour in the fixed $p$ ensembles, i.e. there is no $p - \cal V$ criticality, and hence the  CFT   state dual to a classical charged black hole  cannot be a Van der Waals fluid.  Whether or not  this   phase structure is supported by CFT computations remains an interesting open question.
}
\begin{document}

\maketitle

\section{Introduction}

Black holes, nature's perfect absorbers, are generally believed to   behave as perfect black bodies once   quantum effects are taken into account \cite{hawking1975particle}. The temperature of a black hole depends on its mass, charge, and angular momentum, as well as other parameters pertinent to the physical setup.
In Einstein gravity, the conjugate  thermodynamic entropy is proportional to the horizon area $A$, in notable contrast to an ordinary fluid system, whose entropy is proportional to its volume.  
To be specific, the Hawking temperature and Bekenstein-Hawking entropy of black holes are given by (setting $\hbar=c=k_B=1$) \cite{hawking1975particle,Bekenstein:1973ur}
\be\label{ST}
T=\frac{\kappa}{2\pi}\,,\qquad  S=\frac{A}{4G_N}\,, 
\ee
where $\kappa$ is the surface gravity of the black hole and $G_N$ is the   gravitational Newton  constant. These features indicate that black hole thermodynamics is crucially important in   providing clues about the nature of quantum gravity.  

The thermodynamics of black holes is considerably 
enriched for   asymptotically Anti-de Sitter (AdS) black holes, which have been argued to be equivalent to   thermal states in the dual conformal field theory (CFT)   via the AdS/CFT correspondence \cite{Maldacena:1997re}.  Asymptotically AdS black holes can be in thermal equilibrium with their Hawking radiation and have been shown to
exhibit a plethora of interesting phase behaviour, including  a first-order transition to radiation   \cite{Hawking:1982dh}  corresponding to  confinement/deconfinement of the dual quark gluon plasma~\cite{Witten:1998zw}, Van der Waals   type phase transitions  for charged AdS black holes  \cite{Chamblin:1999tk, Chamblin:1999hg, Cvetic:1999ne,Kubiznak:2012wp}, polymer transitions~\cite{Dolan:2014vba}, reentrant phase transitions
\cite{Altamirano:2013ane,Frassino:2014pha}, triple points
\cite{Altamirano:2013uqa,Wei:2014hba},  and superfluid transitions \cite{Hennigar:2016xwd}. 

The enriched phase behaviour for AdS black holes is due to the presence of a negative cosmological constant~$\Lambda$. In fact, some of the new phase transitions arise by treating the cosmological constant  as an additional thermodynamic variable  for black holes \cite{Kastor:2009wy}.
Recently, Newton's constant $G_N$ has also been added to  the  \emph{extended thermodynamic phase space} as a parameter that can be varied \cite{Kastor:2010gq,Kastor:2014dra,Karch:2015rpa,Sarkar:2020yjs,Visser:2021eqk,Cong:2021fnf}, since it is a coupling constant that can vary in the space of gravitational theories and, in particular, it varies along the renormalization group flow if quantum corrections are included. Moreover, as we will see below, Newton's constant can be useful as a ``bookkeeping device" in finding the correct  thermodynamic interpretation of the extended first law of black holes. 
For  Einstein-Maxwell theory with a negative cosmological constant,   
the \emph{extended first law} of charged AdS black holes in $d+1$ dimensions, including variations of $\Lambda$ and $G_N$,  and the   \emph{generalised Smarr relation}  read, respectively, \cite{Kastor:2009wy,Visser:2021eqk}
 \begin{align} 
 	d M &= \frac{\kappa}{8\pi G_N} dA + \Phi dQ +\frac{\Theta}{8\pi G_N} d \Lambda - \left ( M   - \Phi Q \right)\frac{dG_N}{G_N}\,, \label{eq:extendedfirstlaw11}\\
 		M &=\frac{d-1}{d-2} \frac{\kappa A}{8 \pi G_N} + \Phi Q - \frac{1}{d-2} \frac{\Theta \Lambda}{4 \pi G_N}\,. \label{eq:smarr1111}
 \end{align}
  Here, $M$ is the mass of the black hole, $Q$ is the electric charge and  its conjugate quantity $\Phi$ is the electric  potential. Moreover, $\Theta$ is the quantity conjugate  to the cosmological constant, which can be defined in a geometric way in terms of surface integrals of the Killing potential \cite{Kastor:2009wy} or in terms of the proper volume   weighted locally by the norm of the Killing vector  \cite{Jacobson:2018ahi}.  

The standard thermodynamic interpretation of the negative cosmological constant $\Lambda$ is in terms of a positive \emph{bulk  pressure} \cite{Kastor:2009wy,Dolan:2011xt,Dolan:2010ha,Cvetic:2010jb,Kubiznak:2014zwa} 
\be\label{P}
P=-\frac{\Lambda}{8\pi G_N}\,, \qquad \text{with} \qquad \Lambda=-\frac{d(d-1)}{2 L^2}\,, 
\ee
where   $L$ is the  AdS curvature radius. Assuming Newton's constant is held fixed,   the $\Theta d \Lambda/8 \pi G_N$ term in the extended first law becomes equal to a $VdP$ term, if we identity $V = - \Theta$ as the \emph{thermodynamic volume}. 
The   respective extended first law  \eqref{eq:extendedfirstlaw11} and the  generalised Smarr relation \eqref{eq:smarr1111}  can then be written as  
\ba
d M&=&Td S+\Phi d Q+ Vd P\,,  \label{flaw}\\
M&=&\frac{d-1}{d-2}TS+\Phi Q-\frac{2}{d-2}PV\,.\label{smarr}
\ea

\noindent However,  the bulk pressure interpretation above has some    peculiar  features. First,  the    mass $M$ of the black hole   is  identified with the thermodynamic enthalpy $H$ (rather than internal energy) in this   extended thermodynamic phase space, since the enthalpy satisifes the first law $dH = TdS + VdP$.  This seemingly stands in contrast to the standard identification of the ADM mass with the boundary Hamiltonian generating time translations at asymptotic infinity, which plays the role of an (asymptotic) energy. Second, if we allow for variations of Newton's constant (as a bookkeeping device) then the $\Lambda$ and $G_N$ variations in the extended black hole first law  cannot be combined into a single term $d(\Lambda/ G_N)$. Specfically, the extended first law   \eqref{eq:extendedfirstlaw11} can be rewritten as
\be \label{eq:firstlawfail}
dM = \frac{\kappa}{8\pi } d \left (\frac{A}{G_N} \right)+ \Phi dQ + \frac{\Theta}{8\pi} d \left ( \frac{\Lambda}{G_N} \right) - \left (M - \frac{\kappa A}{8\pi G_N} - \Phi Q -\frac{\Theta \Lambda}{8\pi G_N }  \right ) \frac{dG_N}{G_N}\,.
\ee
The first term on the right can now be identified with the $TdS$ term in the thermodynamic first law, and the third term   can be interpreted as a $VdP$ term. But the final term cannot be put to zero by employing the Smarr relation and it is not clear what its thermodynamic interpretation is. So it seems  that  either   $(P,V)$ are not the right thermodynamic variables for expressing the extended first law, or one has to find another meaningful pair of   thermodynamic variables in addition to $(P,V)$ that accounts for the final term in \eqref{eq:firstlawfail}.\footnote{For instance,     in \cite{Cong:2021fnf}   $dG_N/G_N$  in \eqref{eq:firstlawfail} was replaced by $-\frac{2}{d+1} \frac{dC}{C} - \frac{d-1}{d+1} \frac{dP}{P} $, with $P\sim- \Lambda/ G_N$ being the bulk pressure \eqref{P} and $C\sim L^{d-1}/G_N$   the boundary central charge \eqref{eq:centralcharge}. This yields a   ``mixed'' first law with a new thermodynamic volume and a chemical potential   different from the one in \cite{Visser:2021eqk}. \label{footnote1}}

Moreover,   a  problematic issue of the bulk pressure interpretation is that it does not seem compatible with holography \cite{Johnson:2014yja, Dolan:2014cja, Kastor:2014dra, Zhang:2014uoa, Zhang:2015ova,Karch:2015rpa,Dolan:2016jjc}. In  AdS/CFT  the thermodynamics of AdS black holes can be equivalently described by the dual CFT at finite temperature. Therefore, one would expect that the thermodynamic variables of AdS black holes can be mapped to standard thermodynamic variables in the CFT.  However, the bulk pressure $P$ is not dual to the pressure $p$ of the dual field theory, and the thermodynamic volume $V$ of black holes is not related to the spatial volume  $\cal V$ on which the CFT is formulated \cite{Johnson:2014yja}.  Further,  the generalised Smarr relation should be dual to the thermodynamic Euler relation in the field theory. But the latter relation does not contain any dimension dependent factors whereas the former does, so this raises the question as to how the Smarr relation \eqref{eq:smarr1111} can be mapped to the CFT Euler relation. 

These problems can be resolved by choosing new thermodynamic variables, which at the same time give a correct   CFT   interpretation of extended black hole thermodynamics.  
Several authors \cite{Kastor:2009wy,Dolan:2014cja,Kastor:2014dra,Johnson:2014yja,Karch:2015rpa} have argued that varying the cosmological constant is dual to varying the \emph{central charge} $C$ or the number of colors $N$ in the dual gauge theory. In holographic CFTs dual to Einstein gravity the central charge is related to both the AdS radius and Newton's constant, $C \sim L^{d-1}/G_N$;  hence variations of $C$ in the CFT could in principle lead to variations of both $\Lambda$ and $G_N$ in the gravity theory. Therefore, the   central charge~$C$   and its conjugate \emph{chemical potential} $\mu$ (also called ``color susceptibility" \cite{Karch:2015rpa}) play an essential role in the holographic dual of extended black hole thermodynamics.

Recently, in \cite{Visser:2021eqk} this central charge interpretation has been made more precise by providing an exact match between the extended CFT thermodynamics and the extended gravitational thermodynamics.  
In particular, the extended first law of AdS black holes \eqref{eq:extendedfirstlaw11} was shown to be dual to    the extended CFT first law, where both the field theory pressure and the  central charge  are allowed to vary. By inserting $d\Lambda /\Lambda = - 2 dL/L$ and  using the Smarr relation  \eqref{eq:smarr1111} to  express  $\Theta$ in terms of the other variables, one can write the first law   \eqref{eq:extendedfirstlaw11} as
\be \label{eq:extendedfirstlawnew}
dM  = \frac{\kappa   }{2\pi}   d \left (  \frac{A}{4G_N}    \right  )   + \frac{\Phi}{L} d (Q L )-          \frac{M}{d-1}  \frac{d L^{d-1}}{L^{d-1}}   + \left ( M  -   \frac{\kappa A}{8\pi G_N}     -  \Phi Q \right) \frac{d\! \left ( L^{d-1}/G_N \right) }{ L^{d-1}/G_N} \,.  
\ee
This is a different rewriting of the first law   compared to \eqref{eq:firstlawfail}  in terms of the central charge $C\sim L^{d-1}/G_N$ rather than the bulk pressure $P\sim -\Lambda /G_N$. The advantage of this new rewriting is that all the terms in the first law immediately have a thermodynamic interpretation in the dual CFT, and there are no left-over terms like the $dG_N/G_N$ term in \eqref{eq:firstlawfail},  which do not have a direct thermal interpretation.  In particular, if we insert the standard holographic dictionary for CFTs living on geometries whose  curvature radius coincides with  the AdS   radius $L$  \cite{Karch:2015rpa,Visser:2021eqk},
\be \label{dictionary1}
E=M\, , \qquad \tilde \Phi = \Phi/L \, , \qquad \tilde Q= QL\,, \qquad {\cal V} \sim L^{d-1}\,, \qquad C \sim L^{d-1}/G_N\,,
\ee
and use   \eqref{ST} for the temperature and entropy, then  \eqref{eq:extendedfirstlawnew} turns into the \emph{extended CFT first~law}
\begin{align} 
 	dE &= T dS + \tilde \Phi d \tilde Q - p d{\cal V} + \mu dC\,, \qquad \text{with} \label{cftfirstlawfirst}\\
    \mu &=\frac{1}{C}(E- TS - \tilde \Phi \tilde Q )\,, \quad \text{and} \quad  p = \frac{1}{d-1}\frac{E}{\cal V}\,. \label{mupintro}
\end{align}
Note in \eqref{dictionary1} that both the CFT electric charge and its corresponding conjugate potential are rescaled by the AdS radius \cite{Chamblin:1999tk}.  
In \eqref{mupintro} the formula for the chemical potential $\mu$ is the \emph{Euler equation} in large-$N$ gauge theories in the deconfined phase; it does not contain dimension dependent factors, as should be the case. This Euler equation can be shown to be dual to the generalised Smarr relation  \eqref{eq:smarr1111}  \cite{Karch:2015rpa,Visser:2021eqk}. Further, the field theory pressure $p$ in \eqref{mupintro} satisfies the CFT equation of state in $d$ spacetime dimensions. 

As explained in \cite{Visser:2021eqk}, the holographic dictionary \eqref{dictionary1} can be extended to CFTs for which the curvature radius $R$ is unequal to the AdS length $L,$  which we will review in section \ref{sec:holographic}. In this case there is still a precise match between the CFT first law \eqref{cftfirstlawfirst} and the bulk extended first law~\eqref{eq:extendedfirstlaw11}. The advantage of this  generalised dictionary is that the boundary volume $\cal V$ and the central charge $C$ are now completely independent, since the volume depends on $R$ and the central charge on $L$.   However, one can also set $R=L$ and then all the   results on phase transitions and all the plots presented in the paper remain the same.

The purpose of the present paper is  
to investigate the holographic dual of extended thermodynamics of charged AdS black holes in more detail.  For the three pairs of conjugate quantities    $(\tilde \Phi, \tilde Q)$, $(p, \cal V)$ and $(\mu, C)$, there are in total eight   possible    thermodynamic (grand) canonical ensembles in the CFT. In section \ref{sec:ensembles} we find that three of these ensembles   exhibit interesting phase behaviour or critical phenomena, to wit the ensembles at fixed $(\tilde Q, {\cal V}, C), (\tilde \Phi, {\cal V},C)$ and $(\tilde Q, {\cal V}, \mu)$. We plot the appropriate free energies in these ensembles as a function of the temperature and analyse the relevant phase diagrams. Further, in section~\ref{sec:critistabl} we study the critical behaviour and thermodynamic stability in  detail, and   show that the critical exponents of the critical points in the $\tilde Q-\tilde \Phi$ and $C -\mu$ plane agree with those of mean field theory. Finally, in section  \ref{sec:discussion} we compare our work   with other literature on the holographic dual of extended black hole thermodynamics.

\section{Holographic thermodynamics  of charged AdS black holes}
\label{sec:holographic}
 
The AdS/CFT correspondence relates the thermodynamics of AdS black holes  to the thermodynamics   of the dual CFT \cite{Witten:1998zw}.  
In this section we recap the holographic dictionary for the \emph{extended} thermodynamics of charged AdS black holes, presented in \cite{Visser:2021eqk}. We keep the number of dimensions general in most of the paper, but all plots are made for $d=4$, i.e. the AdS$_5$/CFT$_4$ correspondence.

 \subsection{Extended thermodynamics of charged AdS black holes}
 
  We begin by briefly revisiting the spherically symmetric, charged black holes in asymptotically AdS spacetime \cite{Chamblin:1999tk,Chamblin:1999hg}. This is a solution to Einstein--Maxwell theory with a negative cosmological constant $\Lambda$, whose action in $d+1$ spacetime dimensions reads
 \begin{equation} \label{EMaction}
 	I = \frac{1}{16\pi G_N} \int d^{d+1 } x\sqrt{-g} \left (  R - 2 \Lambda - F^2  \right)
 \end{equation} 
where $F=dA$ is the electromagnetic field strength,  $\Lambda$ is given
in \eqref{P}, where we note that the normalisation $1/16\pi G_N$  of the matter part of the action is   not standard, since it involves Newton's constant.  The metric for a Reissner--Nordstr\"{o}m AdS black hole   in static coordinates is
 \begin{equation} \label{eq:metric1}
 	ds^2= -f(r)dt^2 + \frac{dr^2}{f(r)} + r^2 d \Omega_{d-1}^2\, , 
 \end{equation}
 where $d \Omega_{d-1}^2$ is the metric on the round unit $d-1$ sphere, and the function $f(r)$ is given by
 \begin{equation} \label{eq:blackening}
 	f(r) = 1 + \frac{r^2}{L^2} - \frac{m}{r^{d-2}} + \frac{q^2}{r^{2d-4}} \, . 
 \end{equation}
Here,  $m$ is the mass parameter of the black hole, which  is related to the ADM mass by
  \begin{equation} \label{eq:admmass}
 	M= \frac{ (d-1)\Omega_{d-1}  }{16 \pi G_N}  m \,. 
 \end{equation}
 Further, $q$ is the charge parameter, which is related to the electric charge via
  \begin{equation} \label{eq:electriccharge1}
 	Q = \frac{(d-1)\Omega_{d-1}}{8\pi G_N} \alpha  \,  q \,, \qquad \text{with} \qquad \alpha =\sqrt{ \frac{2(d-2)}{d-1} }\, .
 \end{equation}
 The associated gauge potential is
  \begin{equation}
 	A = \left ( -\frac{1}{\alpha} \frac{q}{r^{d-2}} + \Phi\right) dt\, , 
 \end{equation}
where $\Phi$ is  a constant  that plays the role of the electric potential. 
 Charged black holes have inner horizons and an outer event horizon. We are only concerned in this paper with the outer horizon at $r=r_h$, which is the largest real positive root of $f(r).$ From the condition $f(r_h)=0$ we can solve for the mass parameter in terms of the horizon radius, the AdS radius and the charge parameter
  \begin{equation}
    m=   r_h^{d-2}\left (1 + \frac{r_h^2}{L^2} + \frac{q^2}{r_h^{2d -4}}    \right)\, .
 \end{equation}
 We fix the electric potential such that $A_t(r_h)=0$, i.e.
 \begin{equation} \label{eq:electricpotential1}
 	\Phi = \frac{1}{\alpha}\frac{q}{  r_h^{d-2}} \,. 
 \end{equation}
With this choice $\Phi$ represents the potential difference between the outer horizon and infinity. 
 The black hole parameters are related to each other by the generalised Smarr formula \eqref{eq:smarr1111}  \cite{Kastor:2009wy},
 \begin{equation} \label{eq:smarr}
 	M  =\frac{d-1}{d-2} \frac{\kappa A}{8 \pi G_N} + \Phi Q - \frac{1}{d-2} \frac{\Theta \Lambda}{4 \pi G_N} \, .
 \end{equation}
 Here, $\Theta = -V$ is the quantity conjugate to the cosmological constant $\Lambda$, which can be defined as the   background subtracted Killing volume \cite{Jacobson:2018ahi}
 \begin{equation}
 	\Theta \equiv \int_{\Sigma_{\text{bh}}} |\xi| dV - \int_{\Sigma_{\text{AdS}}} |\xi| dV\,,
 \end{equation}
where $|\xi| = \sqrt{- \xi \cdot \xi}$ is the norm of   the time translation Killing vector $\xi$, which generates the event horizon. We have subtracted the same integral in pure AdS   to cancel the divergence at infinity. Note that the domain of integration $\Sigma_{\text{bh}}$ extends from the horizon to infinity, whereas the domain of integration $\Sigma_{\text{AdS}}$ in the AdS background extends across the entire  spacetime. Moreover, through  Stokes' theorem this definition can be shown to be equivalent to the original definition of $\Theta$ in terms of surface integrals of the Killing potential~\cite{Kastor:2009wy} (see footnote 9 in \cite{Jacobson:2018ahi}). In static coordinates, setting $\xi = \partial_t$, the background subtracted Killing volume is equal to minus  the Euclidean volume excluded by the   black hole, i.e. $\Theta = - \frac{1}{d} \Omega_{d-1} r_h^{d}$ \cite{Kastor:2009wy,Cvetic:2010jb}.
Instead of using the background subraction method one can also employ the counterterm subtraction method to regulate the divergence in the Killing volume \cite{Pedraza:2021cvx}. 

 Further, by using the metric \eqref{eq:metric1} and by computing the surface gravity   defined with respect to the time translation Killing vector $\xi=\partial_t$,  we obtain from \eqref{ST}   the    Hawking temperature  and the Bekenstein--Hawking entropy 
 \begin{equation} \label{eq:entropytemp}
    	T   = \frac{d-2}{4 \pi r_h} \left (1 + \frac{d}{d-2} \frac{r_h^2}{L^2} - \frac{q^2}{r_h^{2d  -4}}\right)\,,\qquad   S  = \frac{\Omega_{d-1} r_h^{d-1}}{4G_N} \, .
 \end{equation}
The generalised Smarr formula  and  the extended first law of AdS black hole mechanics are related by a scaling argument    \cite{Gauntlett:1998fz,Caldarelli:1999xj,Kastor:2009wy}. The black hole parameters scale with the AdS radius as  $  M\sim L^{d-2}$, $A \sim L^{d-1}$, $  Q \sim L^{d-2}$ and $\Lambda\sim L^{-2}$, and with Newton's constant as  $M, Q \sim G_N^{-1}$.   
It can be shown that 
the gravitational first law for charged AdS black holes,  extended to include variations of the theory parameters $G_N$ and $\Lambda$,  is given by
 \begin{equation} \label{eq:extendedfirstlaw}
 	d M = \frac{\kappa}{8\pi G_N} dA + \Phi dQ +\frac{\Theta}{8\pi G_N} d \Lambda - \left ( M   - \Phi Q \right)\frac{dG_N}{G_N}\,.
 \end{equation}
 As mentioned in the introduction, the addition of  the variation of the cosmological constant in the first law is well studied in the literature (see \cite{Kubiznak:2016qmn} for a review). The variation of Newton's constant was considered before in the context of the first law of holographic entanglement entropy in \cite{Kastor:2014dra,Caceres:2016xjz,Rosso:2020zkk} and for the extended first law of AdS black holes in \cite{Kastor:2010gq,Sarkar:2020yjs,Visser:2021eqk,Cong:2021fnf}. 

 \subsection{CFT thermodynamics with a chemical potential for the central charge}
 
In \cite{Karch:2015rpa,Sinamuli:2017rhp, Visser:2021eqk} it was shown that the generalised Smarr formula \eqref{eq:smarr} for AdS black holes is dual to an Euler equation in large-$N$ field theories, which includes a   $\mu C$ term but not a $p {\cal V}$ term. For charged black holes the dual Euler equation takes the following form:
\begin{equation} \label{eq:euler1}
    E = TS + \tilde \Phi \tilde Q + \mu C\,,
\end{equation}
where $\mu$ is the chemical potential associated with the central charge $C,$ and $\tilde \Phi$ and $\tilde Q$ are the CFT electric potential and charge.
Moreover, in  \cite{Visser:2021eqk} the extended first law \eqref{eq:extendedfirstlaw}  was matched with an extended   first law of thermodynamics in the CFT, which   involves both a $\mu dC$ term and a $p d\cal{V}$ term,
\begin{equation} \label{eq:firstlawcft}
   	dE = T dS + \tilde \Phi d \tilde Q - p d{\cal V} + \mu dC\,,
\end{equation}
where $p$ is the field theory pressure and $\cal{V}$ the spatial volume on which the CFT resides. In this subsection we will summarize how this matching works between    the boundary and bulk thermodynamic first laws, for the case where the boundary curvature radius $R$ is unrelated to the bulk curvature radius $L$. We refer to \cite{Karch:2015rpa,Sinamuli:2017rhp,Visser:2021eqk} for a derivation of the generalised   Smarr formula for AdS black holes from the  Euler equation in the  dual CFT. 

Let us begin by explaining the holographic dictionary that is necessary for the match. 
First, in AdS/CFT   the central charge of the CFT is related to the AdS radius and the gravitational coupling constants in the bulk theory. For Einstein gravity this dictionary reads
 \begin{equation} \label{eq:centralcharge}
 	C =   \frac{\Omega_{d-1} L^{d-1}}{16 \pi G_N} \, .
 \end{equation}
In conformal field theory there are several candidates for the central charge in general dimensions. The standard central charges parametrizing the trace anomaly  $\langle {T^\mu}_\mu \rangle$ in a curved background exist only in even dimensions. We normalized the central charge in \eqref{eq:centralcharge} such that it agrees with the coefficient $A$ of the Euler density in the trace anomaly. In $d=2$ our central charge is related to the usual central charge by $C=c/12$; inserting this into \eqref{eq:centralcharge} yields the well-known Brown--Henneaux dictionary $c = 3L/2G_N$ in AdS$_3$/CFT$_2$ \cite{Brown:1986nw}.  Two  other  candidates for a generalised central charge, denoted as $C_T$ and $a^*_d$, are also defined in odd dimensions \cite{Myers:2010xs,Myers:2010tj}. The former central charge 
 $C_T$ is defined as the overall normalization of the two-point function of the CFT stress tensor \cite{Osborn:1993cr}. The latter central charge $a_d^*$ is the universal coefficient in the vacuum entanglement entropy for ball-shaped regions. Now, crucially for CFTs dual to Einstein gravity both $C_T$ and $a_d^*$ scale as $L^{d-1}/G_N$ times a constant in the bulk.  The precise normalization of the central charge is ambiguous, but it can be chosen such that these central charges are equal, $C_T = a_d^*$,   and both  satisfy the holographic dictionary \eqref{eq:centralcharge}~\cite{Hung:2011nu}. Later on, we will see   the normalization of the central charge is   irrelevant for our purpose of    expressing the black hole first law in terms of CFT quantities, since only the combination $dC/C$ appears in the  first law. Therefore,  only the     scaling of the central charge with $L$ and $G_N$, i.e. $C\sim L^{d-1}/G_N$, is important.

Another parameter of the CFT that  appears in the thermodynamic first law, is the spatial volume $\cal{V}$ of the geometry on which the CFT is formulated.  In the literature (see e.g. \cite{Dolan:2016jjc,Karch:2015rpa}) the CFT is often put on a sphere of AdS radius $L$, such that the volume is ${\cal V} =\Omega_{d-1}L^{d-1}$. However, we want to clearly distinguish between the central charge $C$ and the spatial volume ${\cal V}$ of  the CFT. Therefore, we choose the boundary curvature radius $R$ to be different from the bulk curvature radius $L$, and let   the spatial volume be
 \begin{equation}
 \label{vol}
 	{\cal V} = \Omega_{d-1} R^{d-1}\,.
 \end{equation}
 Note  this is the volume of a $(d-1)$-dimensional sphere of radius $R$ in the  CFT boundary geometry $\mathbb R \times S^{d-1}$, which corresponds to an ``area'' in the $(d+1)$-dimensional AdS bulk geometry.
 Technically, this can be realized by choosing a particular conformal frame for the CFT.  In AdS/CFT the dual field theory lives on the conformal boundary of the asymptotically AdS spacetime. More precisely, according to the  Gubser--Klebanov--Polyakov--Witten (GKPW)  prescription \cite{Gubser:1998bc,Witten:1998qj} the CFT metric $ g_{\text{CFT}}$ is identified with the AdS metric $g_{{\text{AdS}}}$ evaluated on the asymptotic boundary up to a Weyl rescaling
 \begin{equation}
     g_{\text{CFT}} =\lim_{r \to \infty}\big( \lambda^2 (x) g_{\text{AdS}}\big)\,,
 \end{equation}
  where $r\to \infty$ corresponds to spatial infinity and $\lambda(x)$ is an arbitrary Weyl factor. As $r \to \infty$ the line element of asymptotically AdS spacetime approaches
  \begin{equation}
 	ds^2 = - \frac{r^2}{L^2 } dt^2 + \frac{L^2}{r^2}dr^2 + r^2 d \Omega_{d-1}^2  \,.
 \end{equation}
For the standard choice of the Weyl factor  $\lambda = L/r $ the  line element of the boundary CFT is $ds^2 = -dt^2 + L^2 d\Omega_{d-2}^2$. In this case the bulk curvature radius $L$ coincides with the boundary curvature radius. However we shall instead take the Weyl factor to be  $\lambda = R/r$ such that the CFT resides on a sphere of constant radius $R$, and the   line element of the CFT becomes
 \begin{equation} \label{eq:cftmetric}
  ds^2 = - \frac{R^2}{L^2} dt^2 + R^2 d\Omega_{d-1}^2\,.
 \end{equation}
 The spatial volume of the boundary sphere is now indeed given by \eqref{vol}. For this choice of CFT metric the holographic dictionary for the   entropy $S$, energy $E$, temperature $T$, electric potential $\tilde \Phi$ and electric charge $\tilde Q$ in the dual field theory  is  \cite{Chamblin:1999tk,Karch:2015rpa,Visser:2021eqk}
 \begin{equation} \label{eq:dictionaryentropyetc}
  S= \frac{A}{4G_N}\,, \qquad 	E = M \frac{L}{R} \,,\qquad  T = \frac{\kappa}{2\pi} \frac{L}{R} \,,\qquad \tilde \Phi =\frac{\Phi}{L}\frac{L}{R}\,,  \qquad \tilde Q = Q L \,.
 \end{equation}
 Note that the factor $L/R$ arises in the dictionary for the energy, temperature and electric potential, since the bulk Schwarzschild time $t$ differs from  the boundary CFT time in \eqref{eq:cftmetric} by a factor $R/L$ \cite{Savonije:2001nd}. Hence, the dictionary for the  energy, temperature and potential differs by a factor $L/R$ from the holographic dictionary in \eqref{dictionary1}. Further, our dictionary here for the charge and potential differs by a factor $\sqrt{G_N}$ from that in our previous work \cite{Cong:2021fnf}, where we used  $\tilde Q  = Q_b L / \sqrt{G_N}$ and $\tilde \Phi  = \Phi_b \sqrt{G_N}/L$ (note we assumed $R=L$). This is because the convention in \cite{Cong:2021fnf} for the bulk charge and potential is different, $Q_b = \sqrt{G_N}Q$ and $\Phi_b = \Phi/ \sqrt{G_N}$, compared to the expressions for $Q$ and $\Phi$ in \eqref{eq:electriccharge1} and \eqref{eq:electricpotential1}, respectively. The difference can be traced back to a different convention for the bulk action: in \cite{Chamblin:1999tk,Karch:2015rpa,Visser:2021eqk} the Einstein--Maxwell action was defined as \eqref{EMaction}, whereas in \cite{Cong:2021fnf} the   bulk action took the standard form  $ I = \frac{1}{16\pi G_N} \int d^{d+1 } x\sqrt{-g} \left (  R - 2 \Lambda - G_N F^2  \right)\, $.    The latter convention leads to another coefficient in the $dG_N$ term in the  bulk first law \eqref{eq:extendedfirstlaw}, since $\Phi d Q + \Phi Q dG_N / G_N =\Phi_b dQ_b + \frac{1}{2}\Phi_b Q_b dG_N/G_N .$  
 
 Furthermore,    in the holographic dictionary  above we assumed the energy of the vacuum state vanishes. However,   the vacuum   energy for a CFT on a sphere is  in fact finite when $d$ is even due to the Casimir effect, and it can be computed in AdS/CFT through  the counterterm subtraction method a.k.a. holographic renormalization \cite{Henningson:1998gx,Balasubramanian:1999re}. The correct dictionary for the renormalized energy that follows from this method is given by  $E =( M + E_0 )L/R$, where the vacuum energy $E_0 \sim C$ is     proportional to the central charge. For the purpose of this paper we can safely ignore the Casimir energy, since it is essentially irrelevant for    the (holographic) thermodynamics, as pointed out in footnote 5 of~\cite{Chamblin:1999hg}. This choice of setting the vacuum energy to zero is also in agreement with other definitions of the asymptotic energy in asymptotically AdS spacetimes \cite{Henneaux:1985tv,Ashtekar:1984zz,Ashtekar:1999jx,Hollands:2005wt}, which are not based on the  counterterm subtraction method. Nevertheless, if it exists,  the Casimir energy does affect the Killing volume~$\Theta$ and the  chemical potential $\mu$; see Appendix E in \cite{Visser:2021eqk} for a renormalized version of the Euler equation.

A crucial step in matching the bulk and boundary first laws is to replace $\Theta$ in the extended first law \eqref{eq:extendedfirstlaw} using the generalised Smarr formula \eqref{eq:smarr}, and to insert $d \Lambda / \Lambda = -2 dL /L$. After a reorganization  we can express the extended first law   in terms of the boundary thermodynamic quantities    \cite{Visser:2021eqk}
 \begin{equation}
\begin{aligned} \label{eq:bulkfirstlaw2}
	d \left (  M \frac{L}{R} \right) &= \frac{\kappa   }{2\pi} \frac{L}{R} d \left (  \frac{A}{4G_N}    \right  )   + \frac{\Phi}{R} d (Q L )-          \frac{M}{d-1} \frac{L}{R} \frac{d R^{d-1}}{R^{d-1}}  \\
	&\qquad + \left ( M \frac{L}{R} -   \frac{\kappa A}{8\pi G_N} \frac{L}{R}    - \frac{\Phi}{R} Q L  \right) \frac{d\! \left ( L^{d-1}/G_N \right) }{ L^{d-1}/G_N} \,.  
\end{aligned}
\end{equation}
The holographic dictionary \eqref{eq:centralcharge}, \eqref{vol} and \eqref{eq:dictionaryentropyetc} then implies that the extended  first law of charged AdS black holes is dual to the following thermodynamic  first law in the CFT
 \begin{equation}
 \label{first law}
 	dE = T dS + \tilde \Phi d \tilde Q - p d {\cal V} + \mu dC\,.
 \end{equation}
By comparing the bulk and boundary first laws, \eqref{eq:bulkfirstlaw2} and \eqref{first law} respectively, we find   the CFT pressure and the chemical potential associated with the central charge are fixed to be
 \begin{equation}
 \label{pmu}
 	p = \frac{1}{d-1} \frac{E}{ {\cal V}}\,, \qquad \qquad \mu  =\frac{1}{C}\left (  E - T S - \tilde \Phi \tilde Q \right)\,.
 \end{equation}
 The first equation is the CFT equation of state in $d$ spacetime dimensions, which is a consequence of the scale invariance of the CFT. The second equation is   the Euler relation \eqref{eq:euler1} for large-$N$ gauge theories with a conserved charge $\tilde Q$, which is equivalent to the fact that the grand canonical free energy is proportional to $C$ (or $N^2$) in the deconfined phase of large-$N$   theories, i.e. $W\equiv E - T S - \tilde \Phi \tilde Q = \mu C$ \cite{Karch:2015rpa}. Notably, in contrast to the generalised Smarr relation~\eqref{eq:smarr}, the Euler equation   does not contain any dimension dependent factors, just like the Euler equation in standard thermodynamics. Moreover, even though the $-pd\cal{V}$ term features in the CFT first law, a $-p{\cal V}$ term is absent in   the large-$N$ Euler equation. This peculiarity arises because the volume does not generically scale with the central charge $C$, like the other thermodynamic quantities in the CFT do, i.e. $E,S, \tilde Q \sim C$. However, in the infinite-volume or infinite-temperature limit $T R \to \infty$ the volume does scale with the central charge, since in that case we have $p {\cal V} = - \mu C$, and the Euler equation reduces to the standard one in flat space $E = TS + \tilde \Phi \tilde Q - p {\cal V}$ (see Appendix C in \cite{Visser:2021eqk}).
 
 The additional term $\mu dC$ in the extended CFT first law \eqref{first law} requires some further discussion. In   $SU(N)$   gauge theories with conformal symmetry like $\mathcal N=4$ supersymmetric Yang-Mills theory the central charge is associated with the rank of the gauge group, $C\sim N^2.$ Changing the rank of the gauge group is tantamount to changing the original theory. Variations of the rank of the gauge group can thus be viewed as moving within the space of theories and changing the number of degrees of freedom $N^2$. We are investigating in this paper how various physical quantities change as the central charge varies. However, since  $N$ is an integer it can not be varied in a continuous way. Nevertheless, in the large-$N$ limit, which is relevant for holography, we   still have $\Delta N/ N \ll 1$, and hence $d C/C \ll 1.$ Since $\mu \sim 1/C$ the combination $\Delta C/C$ appears precisely in the first law, and therefore the $\mu dC$ term in the first law only makes sense for large-$N$ field theories.\footnote{We thank an anonymous referee for pointing this out.} %Secondly, below we will also consider ensembles where $\mu$ is kept fixed

   Finally, we give explicit expressions for the thermodynamic variables of CFT thermal states dual to charged AdS black holes. 
  Following  \cite{Dolan:2016jjc},  we find it convenient to introduce the  dimensionless parameters
   \begin{equation}
 	x \equiv \frac{r_h}{L} \,,\qquad   y \equiv \frac{q}{L^{d-2}}\,.
 \end{equation}
The gravitational entropy \eqref{eq:entropytemp}, electric charge \eqref{eq:electriccharge1} and potential \eqref{eq:electricpotential1} of AdS black holes can be mapped to CFT  variables using the  holographic dictionary    \eqref{eq:dictionaryentropyetc}:
\begin{equation}
 \label{SQphi}
 	S = 4 \pi C x^{d-1}, \qquad \tilde Q = 2 \alpha (d-1)  C  y , \qquad \tilde \Phi = \frac{1}{\alpha R} \frac{y}{  x^{d-2}} \, , 
 \end{equation}
 where we recall $R$ is the boundary curvature radius,   and $C$ is the CFT central charge \eqref{eq:centralcharge}.
Further, 
 the ADM  energy \eqref{eq:admmass} and the Hawking temperature \eqref{eq:entropytemp} can also be related to the   CFT energy and temperature, respectively,
 \begin{equation} \label{energytempcft}
E = \frac{d-1}{R} C x^{d-2} \left ( 1 + x^2 + \frac{y^2}{x^{2d-4}} \right), \qquad T = \frac{d-2}{4 \pi R} \frac{1}{x} \left ( 1 + \frac{d}{d-2} x^2 - \frac{y^2}{x^{2d-4}} \right)\,,
 \end{equation}
  where \eqref{SQphi} can be used to express $x$ and $y$ in terms of the CFT variables $S, \tilde Q$, $\tilde \Phi$, $C$ and~$R.$ Moreover,  the chemical potential \eqref{pmu}  can be expressed in terms of $x, y$ and $R$ as
 \begin{equation}
 \label{mu}
 	\mu = \frac{x^{d-2}}{R} \left ( 1 - x^2 - \frac{y^2}{x^{2d -4}}\right)\,.
 \end{equation}
We point out that the $1/R$  dependence   in equations \eqref{SQphi}-\eqref{mu} is fixed by the scale invariance of the CFT; the scale invariant combinations are $\tilde \Phi R,$ $E R $, $TR$ and $\mu R$.

 \section{Thermodynamic ensembles in the CFT}
 \label{sec:ensembles}
 
 Using the holographic dictionary of the previous section we want to study different (grand) canonical  thermodynamic ensembles in the CFT.   Besides the pair $(T,S)$ there are three pairs of conjugate thermodynamic variables in the CFT description of charged AdS black holes,  namely $(\tilde \Phi, \tilde Q)$, $(p,{\cal V})$ and $(\mu, C)$. For the study of thermal behavior this yields $2^3=8$ (grand) canonical ensembles. 
  
 Five of these ensembles 
 exhibit no interesting phase behaviour.   The relevant grand canonical potentials or free energies for these   ensembles are 
  \begin{equation}
     \begin{aligned} \label{freeenergies1}
    &\text{fixed} \quad  (\tilde Q, p,C  ): \qquad &&F_1 \equiv E - TS +p{\cal V}= \tilde \Phi \tilde Q +\mu C + p{\cal V} \,,\\
      &\text{fixed} \quad  ( \tilde Q, p,\mu  )\,: \qquad &&F_2 \equiv E- TS+p{\cal V} - \mu C = \tilde \Phi \tilde Q + p{\cal V}\,,\\
       &\text{fixed} \quad  (\tilde \Phi,p,\mu)\,: \qquad &&F_3 \equiv E - TS- \tilde \Phi \tilde Q +p{\cal V} - \mu C=p{\cal V}\,,\\
     &\text{fixed} \quad  (\tilde \Phi,p,C): \qquad &&F_4 \equiv E-TS - \tilde \Phi \tilde Q +p{\cal V}= \mu C + p{\cal V}\,,\\
     &\text{fixed} \quad  (\tilde \Phi, {\cal V},\mu): \qquad &&F_5 \equiv E-TS - \tilde \Phi \tilde Q - \mu C=0\,. 
     \end{aligned}
 \end{equation}
 We did find  phase transitions or  critical phenomena  
 for the  remaining three ensembles at fixed $(\tilde Q, {\cal V}, C), (\tilde \Phi, {\cal V},C)$ and $(\tilde Q, {\cal V}, \mu)$, whose free energies we denote as $F, W$ and $ G$,   respectively,
   \begin{equation}
     \begin{aligned}  \label{freeenergies2}
    &\text{fixed} \quad  (\tilde Q, {\cal V}, C): \qquad &&F \equiv E - TS  = \tilde \Phi \tilde Q +\mu C \,,\\
      &\text{fixed} \quad  (\tilde \Phi, {\cal V},C)\,: \qquad &&W \equiv E- TS - \tilde \Phi \tilde Q =\mu C\,,\\
       &\text{fixed} \quad  (\tilde Q, {\cal V}, \mu)\,: \qquad &&G \equiv E - TS-     \mu C=\tilde \Phi \tilde Q\,. 
     \end{aligned}
 \end{equation}
 In the next three subsections \ref{sec:firstensemble}$-$\ref{sec:thirdensemble} we will study the phase behaviour of these three ensembles   by looking at how the appropriate free energies behave as functions of temperature. In the last subsection \ref{sec:otherensembles} we will briefly mention our findings in the other five ensembles.  
 
 \subsection{Ensemble at fixed ($\tilde Q, {\cal V}, C$)}
 \label{sec:firstensemble}
 
 \begin{figure}
    \centering
    \includegraphics[scale=0.75]{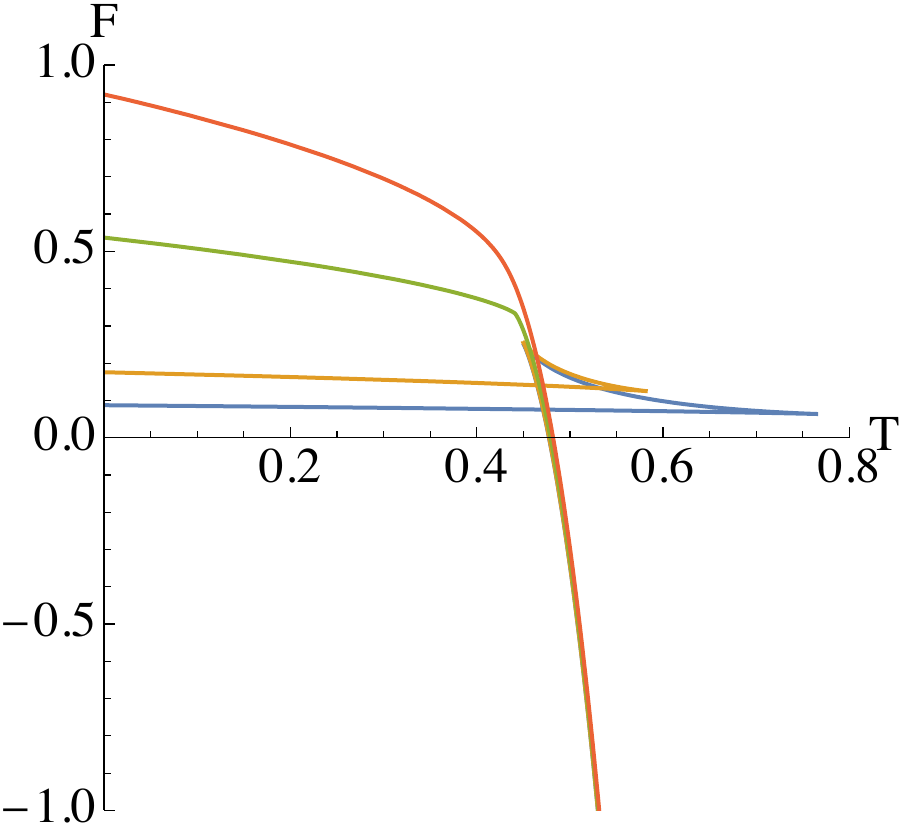} \hspace{1 cm}
    \includegraphics[scale=0.75]{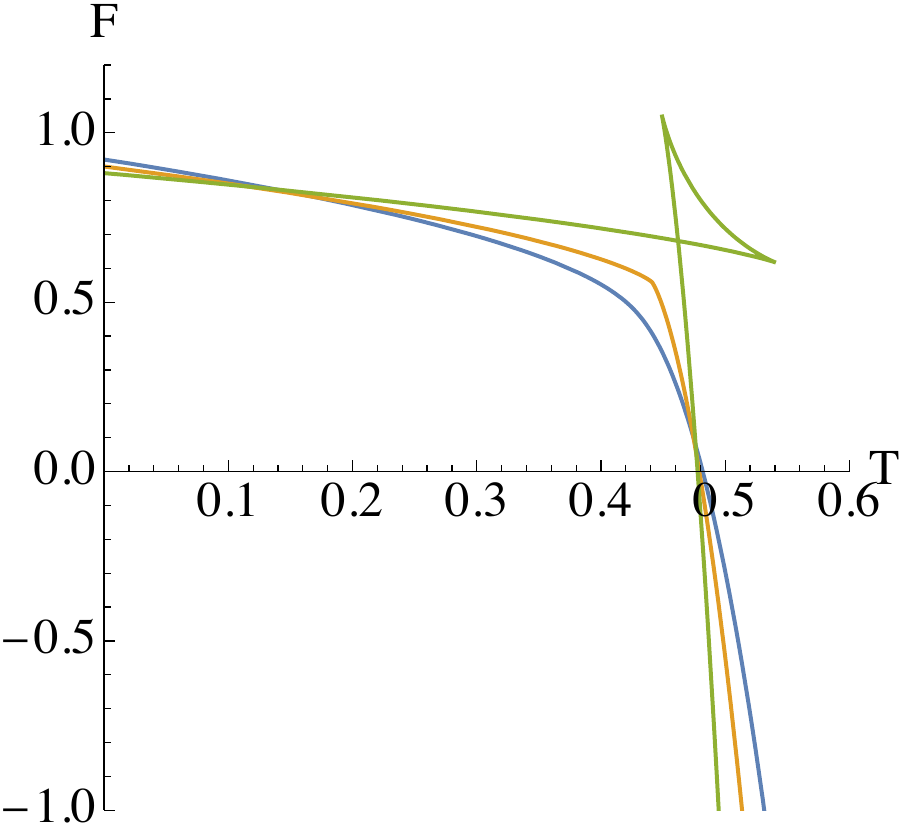}
    \caption{Free energy $F$ vs. temperature $T$ plot   in $d=4$ for the fixed $(\tilde Q,{\cal V},C)$ ensemble. \textbf{Left}:   we plot different values of $\tilde Q$ for fixed $R$ and $C$, the  parameters   are   $R=1$, $C=1$ and $\tilde Q=0.1,0.2, 4/3 \sqrt{5}, 1$ (blue, orange, green, red). For $Q< Q_{crit}$ (blue, orange) the free energy displays     ``swallowtail'' behaviour and a first-order phase transition occurs between two thermodynamically stable branches. The ``horizontal'' branch has low entropy, while the ``vertical'' branch has high entropy. The intermediate branch that connects these two branches has negative heat capacity and is hence unstable.  
    For $Q=Q_{crit}$ (green) there is a second-order phase transition,   and for $Q > Q_{crit}$ (red) there are no phase transitions. 
    \textbf{Right}: We plot different values of $C$ for fixed $\tilde Q$ and $ R$, the  parameters   are  $R=1$, $\tilde Q=1$ and $C=1, 3\sqrt{5}/4, 4$ (blue, orange, green). As in the charged case, the plot exhibits different behaviours below and above the critical value, this time, a critical central charge. The swallowtail and first-order phase transitions occur for $C>C_{crit}$ (green). This transition becomes of second order at $C=C_{crit}$ (orange) and for $C<C_{crit}$ only a single phase exists, as implied by the smooth single-valued curve (blue).   The two apparent triple intersections of the  blue, orange and green curves is a consequence of plotting resolution; there are no triple intersections.  
    }
    \label{fig:qvcFT}
\end{figure}

  In the    canonical ensemble   we fix the electric charge $\tilde Q$, the spatial volume ${\cal V}$ and the central charge $C$.  The   thermodynamic potential in this ensemble is the  Helmholtz free energy  
 \begin{align}
 \label{F1}
 	F  \equiv E - T S 
 	 = C \frac{x^{d-2}}{R} \left( 1 - x^2 + (2d-3) \frac{y^2}{x^{2d-4}}\right)\,.  
 \end{align}
 Indeed by the CFT first law  \eqref{first law}, the differential of $F$ satisfies 
 \begin{equation}
     dF = dE-TdS-SdT = -SdT+\tilde \Phi d\tilde Q-pd{\cal V}+\mu dC\,.
 \end{equation}
 Hence, $F$ is    stationary at   fixed   $(T,\tilde Q,{\cal V},C)$, and is therefore the right free energy in this case. This ensemble is equivalent to the fixed charge ensemble analysed in  \cite{Chamblin:1999tk,Chamblin:1999hg}, but we clarify that in  the dual CFT description  implicitly  ${\cal V}$ and   $C$ are also kept~fixed. 
 
 Let us now study how the free energy $F$ behaves as a function of the temperature $T$ for different fixed ($\tilde Q,{\cal V},C$) values. For this purpose, it is practical to consider $F$ and $T$ as functions of ($\tilde Q,R,C,x$), where we note that fixing the radius $R$ is identical to fixing the volume ${\cal V}$ due to \eqref{vol}. This is done by replacing the parameter $y$ in \eqref{energytempcft} and \eqref{F1} by the electric charge and central charge, using the relation   $\tilde Q = 2 \alpha (d-1) C y$. We then obtain for the free energy and temperature, respectively,
 \begin{equation}
 \begin{aligned}
 	F  &=   C \frac{x^{d-2}}{R} \left( 1 - x^2 +  \frac{2d-3}{4 \alpha^2 (d-1)^2 C^2}  \frac{{\tilde Q}^2}{x^{2d-4}}\right)\,,\\
 	T &= \frac{d-2}{4 \pi R} \frac{1}{x} \left ( 1 + \frac{d}{d-2} x^2 -  \frac{2d-3}{4 \alpha^2 (d-1)^2 C^2}  \frac{\tilde Q^2 }{x^{2d-4}} \right)\,.
 \end{aligned}
 \end{equation}
 This allows us to plot $F(T)$ parametrically using $x$ as parameter for fixed values of ($\tilde Q,R,C$), which we depict in Figure \ref{fig:qvcFT}. The dependence on the radius $R$ is trivially fixed by scale invariance, hence  plots of $F(T)$ for different values of $R$ are only rescalings of each other. It does matter, however,  whether we plot $F(T)$ for different values of $\tilde Q$, while keeping $C$ fixed, or for different values of $C$, while keeping $\tilde Q$ fixed.  In both cases the free energy displays ``swallowtail'' behaviour connecting three different branches, but in the former case this only occurs   for $Q< Q_{crit}$ and in the latter case for $C > C_{crit}$, where $Q_{crit} $ and $C_{crit}$ are  critical values whose ratio we compute below in   equation \eqref{Ccrit}.  In Figure \ref{fig:qvcFT}  the former case is shown on the left and the latter case is shown on the right. We will now discuss these two different plots in more detail.   
  
 On the left in Figure  \ref{fig:qvcFT} we show the free energy as a function of the temperature for $Q < Q_{crit}$ (blue, orange), $Q=Q_{crit}$ (green) and $Q > Q_{crit}$ (red) while keeping $C$ and $R$ fixed. 
 The free energy displays a ``swallowtail'' shape for $\tilde Q<\tilde Q_{crit}$, a kink when $\tilde Q=\tilde Q_{crit}$, and a smooth monotonic curve for $\tilde Q>\tilde Q_{crit}$. For each of the curves, starting from the point on the curve where $T=0$, the value of $x$ along the curves increases as $T$ increases. From the formula $S=4 \pi C x^{d-1}$ for the CFT entropy, we see that black holes with small $x\equiv r_h /L$  are dual to CFT thermal states with small $S/C$, i.e.  states with \emph{low entropy per degree of freedom}. On the swallowtail curve (e.g. blue), this low-entropy state (from here on we drop the phrase ``per degree of freedom'' to avoid clutter) is the only available state near $T=0$ on this curve and thus has initially the lowest free energy $F$. It continues to have the lowest free energy as $T$ increases until the self-intersection point of the curve. Beyond this point, the CFT state with \textit{high entropy per degree of freedom}, corresponding to large $x$ black holes,  lying along the ``vertical'' branch of the curve, becomes the state with lowest free energy $F$  and hence dominates the canonical ensemble. A first-order phase transition thus takes place between low- and high-entropy states at the self-intersection temperature for each value of $\tilde Q < \tilde Q_{crit}$. As we increase $\tilde Q$, the temperature at which the   phase transition occurs increases while the swallowtail shrinks in size until $\tilde Q=\tilde Q_{crit}$, where it becomes just a kink in the curve. The phase transition between low- and high-entropy states becomes second order at this $(\tilde Q_{crit}, T_{crit})$ critical point, which depends on the value of $C$. This $F-T$ behaviour of the CFT is commensurate with the canonical ensemble for AdS black holes at fixed charge~\cite{Chamblin:1999tk}.
 
However unlike \cite{Chamblin:1999tk}, we also consider  variations of $C$ while keeping $\tilde Q$ fixed. On the right in Figure \ref{fig:qvcFT} we show the plots of $F(T)$ for three representative values of $C$ while keeping $\tilde Q$ and $R$ fixed: $C<C_{crit}$ (blue), $C=C_{crit}$ (orange) and $C>C_{crit}$ (green). Qualitatively, this plot conveys a phase behaviour similar to that on the left figure. That is, the CFT displays a first-order phase transition between states with low- and high-entropy per degree of freedom for $C>C_{crit}$, a second-order phase transition at $C_{crit}$, and a single phase for $C<C_{crit}$. As we decrease $C$, the temperature at which the first-order phase transition occurs decreases  until the critical central charge   is reached, where the phase transition becomes second order. Comparing   the left and right plots in Figure \ref{fig:qvcFT}, it is noteworthy that the first-order phase transitions occur for large central charge $C>C_{crit}$ on the right, but for small electric charge $\tilde Q< Q_{crit}$   on the left. Moreover, on the right the value of the free energy at which the first-order phase transition occurs decreases  as $C$ decreases; in contrast, on the left the free energy of the first-order phase   transition  increases  as $\tilde Q$ decreases.    These results on central charge criticality are consistent with recent findings from a bulk perspective \cite{Cong:2021fnf}.

 \subsection{Ensemble at fixed ($\tilde\Phi, {\cal V}, C$)}
 
  \begin{figure}
     \centering
     \includegraphics[scale=0.75]{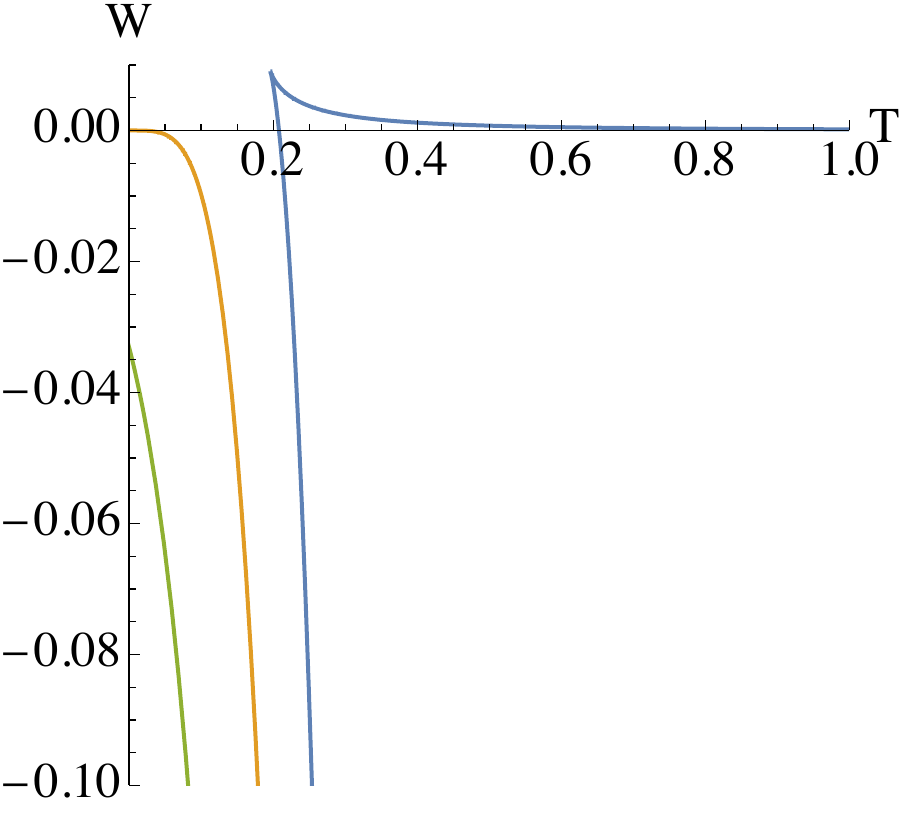} \hspace{1cm}
     \includegraphics[scale=0.75]{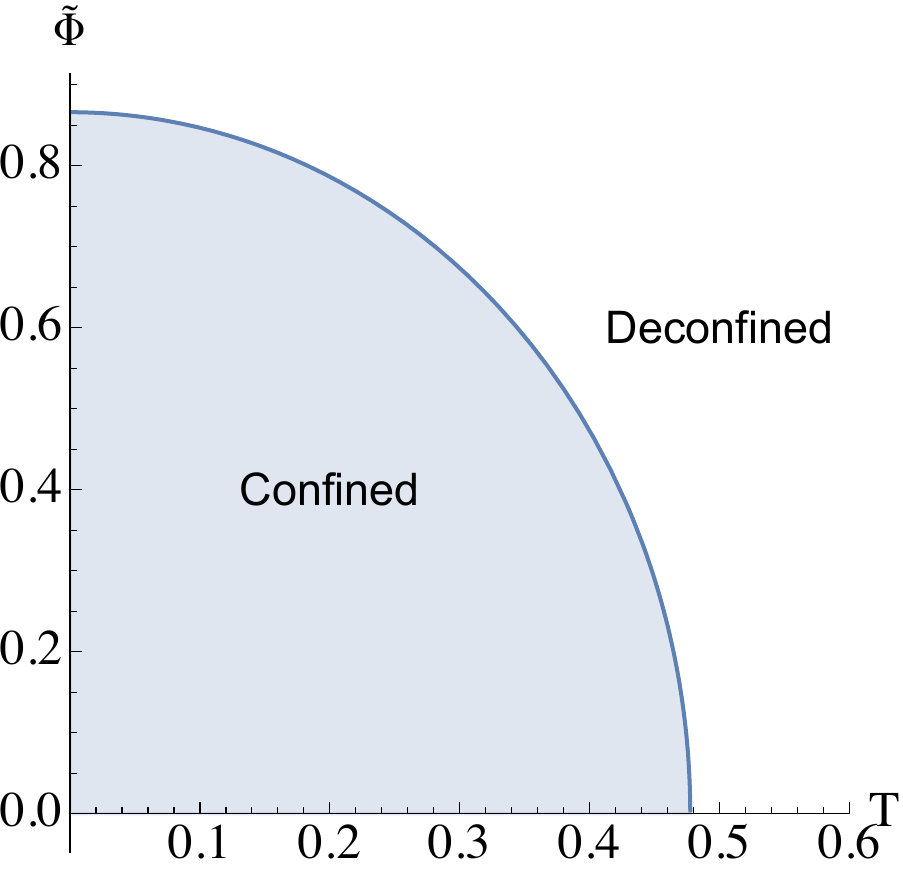}
     \caption{Free energy $W$ vs. temperature $T$ plot and phase diagram for the fixed $(\tilde \Phi,{\cal V},C)$ ensemble in $d=4$.  \textbf{Left}: $W - T$ plot for the parameters  $R=1$, $C=1$ and $\tilde\Phi=0.9\Phi_c$ (blue), $\tilde\Phi=\Phi_c =  \sqrt{3}/2$ (orange) and $\tilde\Phi=1.1\Phi_c$ (green). For $\tilde\Phi<\Phi_c$  the free energy curve   consists of two branches ending in a cusp,  the upper branch corresponds to a low-entropy state and the lower branch to a high-entropy state. At $W=0$ there is a   first-order phase transition between the high-entropy ``deconfined'' state and a ``confined'' state,  dual to a generalised Hawking-Page   transition between a large black hole and thermal AdS, which depends on the value of $\tilde \Phi$ here.
     For $\Phi \ge \Phi_c$  the curve  lies below or at  the $W=0$ axis and no phase transition takes place. \textbf{Right}: The  $\tilde \Phi-T$ phase diagram for $R=C=1$ with a coexistence curve  representing a line of (de)confinement phase transitions in the CFT.
At $\tilde \Phi=0$ the transition occurs at the Hawking-Page temperature $T=T_c=3/2\pi$, and at $T=0$ it happens at $\tilde \Phi =\tilde \Phi_c.$ } 
     \label{fig:WTplot}
 \end{figure} 
 
 If we fix the potential $\tilde \Phi$, instead of the charge $\tilde Q$, then we are in the grand canonical ensemble at fixed ($\tilde\Phi, {\cal V}, C$). The thermodynamic potential of this ensemble is the    Gibbs  free energy 
  \begin{equation} \label{eq:freeenergy2}
 	W \equiv E - TS - \tilde \Phi \tilde Q = \mu C =C \frac{x^{d-2}}{R} \left ( 1 - x^2 - \frac{y^2}{x^{2d -4}}\right)  \,,
 \end{equation}
 where we inserted the Euler equation \eqref{eq:euler1}, which implies the free energy $W$ is just proportional to $\mu$, and  in the last equality we used \eqref{mu}.  This ensemble is equivalent to the fixed potential ensemble considered in \cite{Chamblin:1999tk,Chamblin:1999hg}, but here we give a more precise interpretation in terms of the dual CFT (see also~\cite{Dolan:2016jjc}). 
 
 Using   $\tilde \Phi = y/ \alpha R x^{d-2}$ we can rewrite the free energy in terms of the electric potential
 \begin{equation}  
 	W   =C \frac{x^{d-2}}{R} \left ( 1 - x^2 -  \alpha^2 R^2 \tilde \Phi^2\right)  \,.
 \end{equation}
Similarly, the temperature can be expressed as
 \begin{equation}
  T = \frac{d-2}{4 \pi R} \frac{1}{x} \left ( 1 + \frac{d}{d-2} x^2 - \alpha^2 R^2 \tilde \Phi^2 \right)\,. 
 \end{equation}
This allows us to plot the free energy $W(T)$ parametrically using $x$ as parameter for fixed $(\tilde \Phi, R, C)$, as shown in Figure \ref{fig:WTplot} on the left.  The $W-T$ plot displays different behaviour above and below a certain critical potential $\tilde \Phi_c$, which we compute below in \eqref{linefixedphi}.    On the one hand, for $\tilde\Phi\geq \tilde \Phi_c $ (orange, green curves) the free energy  is single valued as a function of the temperature, with $W\leq 0$ and the curve cuts the $W$-axis. On the other hand, for $\tilde\Phi<\tilde \Phi_c$ (blue curve) the free energy curve consists of an upper and a lower branch that meet at a cusp, corresponding to low-entropy (small black holes) and high-entropy (large black holes) states, respectively. The upper branch  has negative specific heat and is hence thermodynamically unstable, while the lower branch  has positive specific heat and is thus a stable solution. The temperature attains a minimum as a function of $x$ at the cusp, i.e.
\begin{equation}
\begin{aligned}
   &\left ( \frac{\partial T}{\partial x} \right)_{\tilde \Phi}=0 \quad \text{at}\quad x_{cusp} = \sqrt{\frac{d-2}{d}(1-\alpha^2  R^2\tilde \Phi^2)} \,,\\
   &\,\,\,\text{hence} \qquad T_{cusp} = \frac{1}{2\pi R}\sqrt{d(d-2)(1-\alpha^2  R^2\tilde \Phi^2)}\,. \label{cusp}
   \end{aligned}
\end{equation}
We see from the blue curve on the left of Figure \ref{fig:WTplot} that the free energy of the lower branch switches sign at $W=0$,   signalling a first-order phase transition. The large-entropy ``deconfined'' state  dominates the ensemble when $W<0$, while the ``confined'' state is thermodynamically preferred when $W>0$. This (de)confinement phase transition is dual to a generalised Hawking-Page   phase transition   between large AdS black holes and the AdS spacetime with thermal radiation \cite{Witten:1998zw,Hawking:1982dh}. Even though the Hawking-Page transition was originally discovered for   AdS-Schwarzschild black holes \cite{Hawking:1982dh}, it  also  exists for   charged AdS black holes, where the transition depends on the value of the electric potential \cite{Chamblin:1999tk}. This generalised Hawking-Page transition even exists for Lifshitz black holes for the values $1 \le z \le 2$ where $z$ is the dynamical Lifshitz exponent \cite{Tarrio:2011de} and for hyperscaling violating black holes for any hyperscaling violation parameter $\theta$ \cite{Pedraza:2018eey}. Thus, there exists an entire line in the $\tilde \Phi - T$ plane along which   first-order phase transitions between the confined and deconfined phase occur.

This line of first-order phase transitions     can be computed analytically. By 
setting $W=0$ and eliminating $x$ in favour of the temperature, we can obtain the following expression for the coexistence line
\begin{align}
 \tilde \Phi &= \frac{\tilde \Phi_c}{T_c} \sqrt{T_c^2 - T^2} \,, \qquad \text{with} \qquad T_c = \frac{d-1}{2\pi R} \,,\qquad \tilde \Phi_c =\frac{1}{\alpha R} =\frac{1}{R} \sqrt{\frac{d-1}{2d-4} }\,. \label{linefixedphi}
 \end{align}
  We plotted  the coexistence line in the $\tilde \Phi - T$ plane in Figure \ref{fig:WTplot} on the right. Notice  at $T=0$ the phase transition  occurs at $\tilde \Phi = \tilde \Phi_c$, and at $\tilde \Phi=0$ the   transition is equivalent to the standard Hawking-Page phase transition in the bulk  at temperature $T=T_c.$ The  values for $T_c$ and $\tilde \Phi_c$ in \eqref{linefixedphi} are consistent with the expressions found in the bulk in \cite{Chamblin:1999tk}. Even though the phase behaviour is essentially the same as for the  fixed potential ensemble in the bulk, our analysis makes  sure  that the volume $\mathcal V$ and central charge $C$  are also explicitly kept fixed in the corresponding fixed potential ensemble  %also need to be kept fixed in the fixed potential ensemble 
  on the boundary.

 \subsection{Ensemble at fixed ($\tilde Q, {\cal V}, \mu$)}
 \label{sec:thirdensemble}
 
 \begin{figure}
     \centering %0.7 and 0.325
     \includegraphics[scale=0.75]{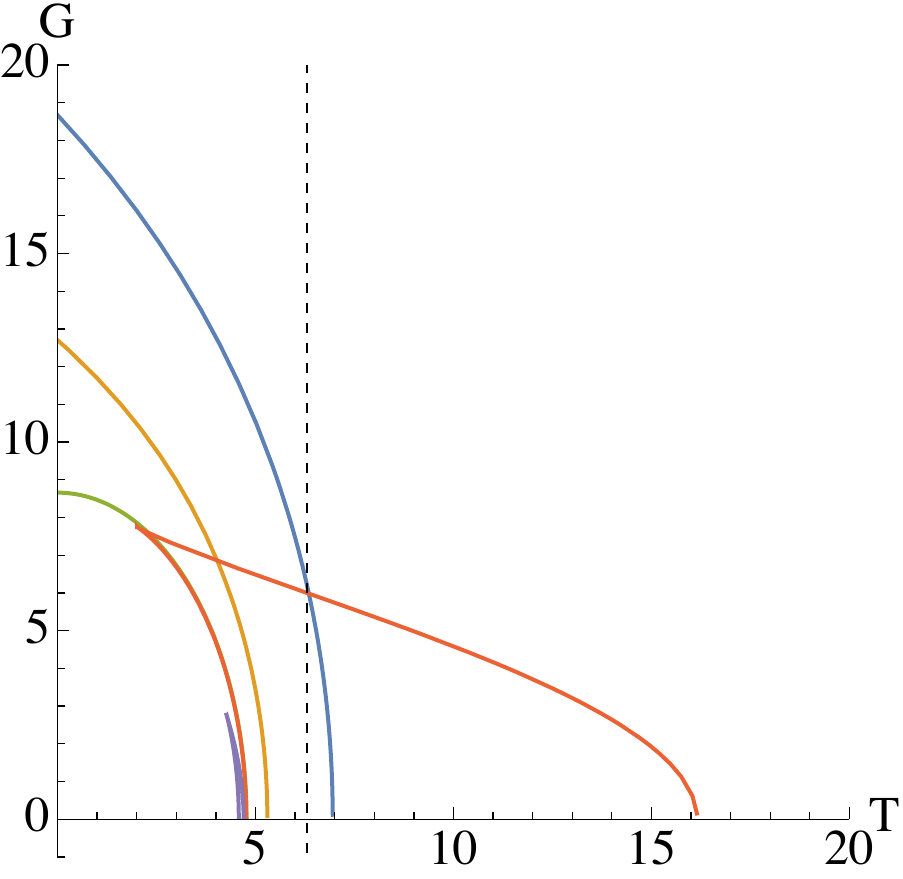} \hspace{1cm}
     \includegraphics[scale=0.335]{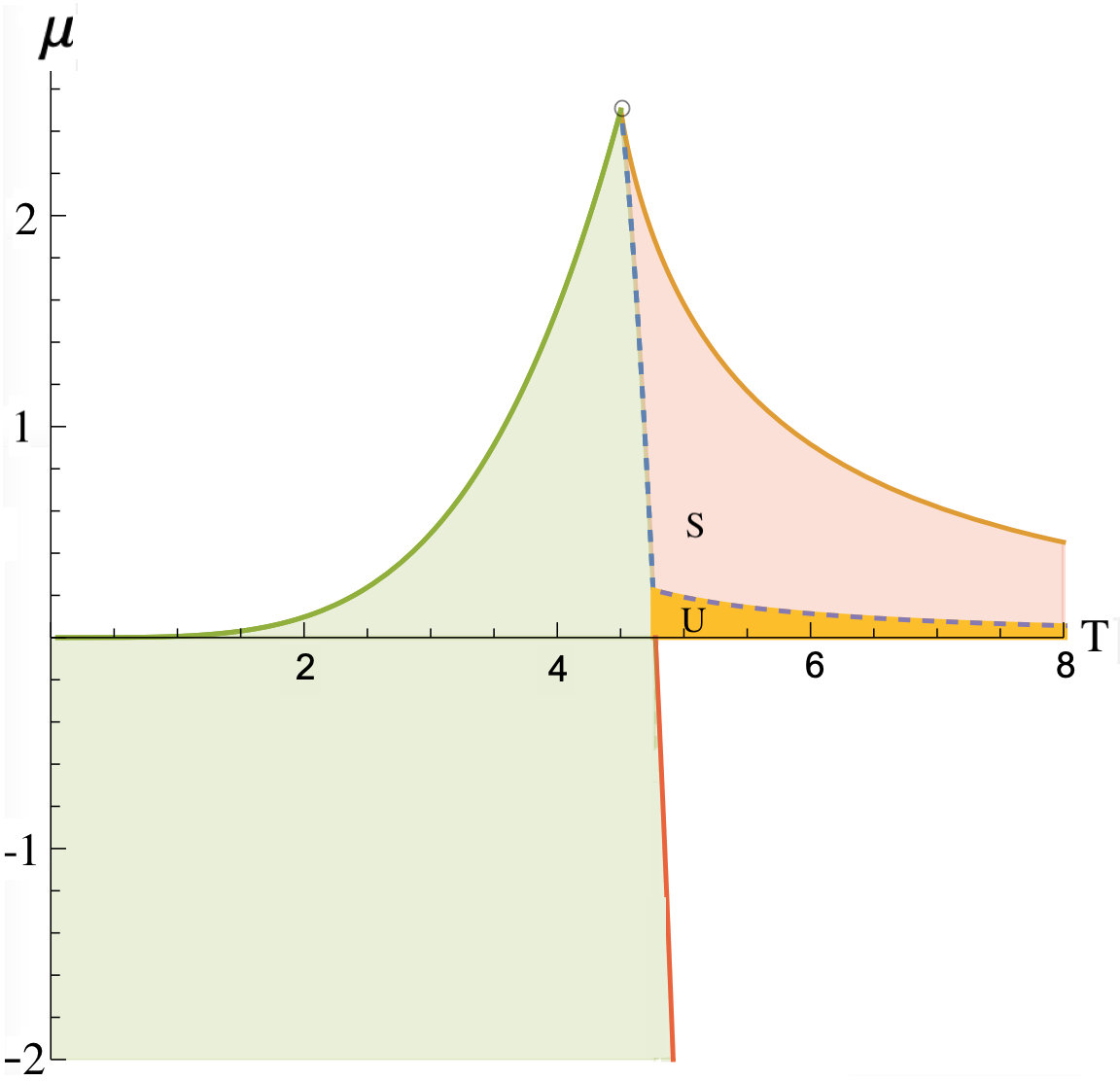}
     \caption{Free energy $G$ vs. temperature $T$ plot and phase diagram for the fixed $(\tilde Q,{\cal V},\mu)$ ensemble in $d=4$. In both figures we set $R=0.1$, $\tilde Q=1$.
      \textbf{Left}: The $G(T)$ plot for the values $\mu=-60,-10,0,1/10,2$ (blue, orange, green, red, purple). For $\mu\leq 0$  there is only a single stable phase. For $\mu>0$ (e.g. red curve) the free energy curve consists of two branches, which meet    at a cusp $T=T_{0}$ \eqref{cusptemp} and   cut the $G=0$ line at two temperatures $T_1\leq T_2$ \eqref{TLR}. The upper  branch corresponds to a low-entropy state and the lower  branch to a high-entropy state. For $T_{0}<T<T_1$  the high-entropy state is the thermodynamically preferred phase. At $T=T_1$, the high-entropy branch terminates and the CFT undergoes a zeroth-order phase transition to the low-entropy CFT phase when the temperature is increased. The low-entropy branch has positive heat capacity at $T=T_2$, but it becomes negative at some intermediate temperature $T_{int}$ indicated by the black dashed line for the red curve. \textbf{Right}: The  $\mu-T$ phase diagram showing demarcations between the different phases: the green shaded region bounded by the green curve is the high-entropy phase, which has positive heat capacity, the regions on the right above the $T$-axis correspond to  the low-entropy phase. The latter is further divided into a stable (``S'') phase with positive heat capacity and an unstable (``U'') phase with negative heat capacity. White regions indicate that no solution exists. }
     \label{fig:GTplot}
 \end{figure}
 
In the previous two ensembles we fixed the central charge $C$, which roughly corresponds to fixing the number of field degrees of freedom in the CFT or     $N^2$ in a large-$N$ $SU(N)$ gauge theory. Here we consider what happens if one fixes the associated chemical potential $\mu$ instead.   The appropriate free energy for this grand canonical ensemble at fixed $(\tilde Q, {\cal V},\mu)$ is
\begin{equation} \label{eq:freeenergy3}
	G \equiv E - T S - \mu C = \tilde \Phi \tilde Q\,,
\end{equation}
where we inserted the Euler equation \eqref{eq:euler1} in the last equality. 
 The differential of $G$ is
\begin{equation}
    dG = dE - TdS - S dT - \mu dC - C d \mu = - S dT + \tilde \Phi d \tilde Q - p d {\cal V} - C d \mu\,,
\end{equation}
so the free energy $G$ is indeed stationary at fixed $(T, \tilde Q, {\cal V}, \mu).$ This ensemble has not been considered before in the black hole thermodynamics literature. We note that fixing the chemical potential instead of the  number of degrees of freedom is very natural in   thermodynamics.  However, in the CFT this means that we are allowed to vary the central charge, which is only possible if we consider a family of holographic CFTs with different central charges. In the gravity theory this corresponds to allowing for variations of $\Lambda$ and $G_N$.  

We can express the free energy and temperature as functions  $G  = G(  \tilde Q, R, \mu,x)$ and  $T = T( R, \mu,x)$ by replacing the potential $\tilde \Phi$ with the chemical potential $\mu$ using equations \eqref{mu} and \eqref{SQphi}
\begin{align}
 G    &= \frac{|\tilde Q|}{\alpha  R  } \sqrt{1-x^{ 2}- \frac{R\mu}{x^{d-2}} } \,,	\label{TG}\\ \qquad 	T &= \frac{d-2}{4 \pi R} \left (\frac{R\mu}{x^{d-1}}  + \frac{2(d-1)}{d-2}x \right) \,. \label{TG2}
\end{align}
In the formula for $G$ the  absolute value of $\tilde Q$ appears, since strictly speaking the free energy   cannot become negative. This can be easily seen by expressing the free energy in terms of the bulk electric potential and charge, i.e. $G=\frac{L}{R} \Phi Q=\frac{L}{R} \frac{(d-1)\Omega_{d-1}}{8\pi G_N} \frac{q^2}{r_h^{d-2}}$, hence the free energy is always greater than or equal to zero:   $G  \ge 0$.

 With these expressions \eqref{TG} and \eqref{TG2} we can plot $G(T)$ parametrically for fixed values of $(\tilde Q, R, \mu)$; see the left plot of Figure \ref{fig:GTplot}. We note that  the $G(T)$ diagram undergoes a qualitative change at $\mu = 0$ (the green curve), in which case the temperature is given by $T (\mu=0) = \frac{d-1}{2 \pi R} x  $. For $\mu\le 0$ (blue, orange and green curves) the free energy as a function of the temperature is single valued, corresponding to the existence of a single CFT phase that is stable under thermal fluctuations. For $0<\mu<\mu_{coin}$ (red and purple curve) the free energy  curve consists of two branches, cutting the $G=0$ line at two temperatures $T_1$ and $T_2$ with $T_2\ge T_1$,  corresponding to the two positive roots of the function  $f(x)=1-x^{ 2}- \frac{R\mu}{x^{d-2}}$ appearing in expression   \eqref{TG}. For $d=4$ we can find the analytic expressions 
\begin{equation} \label{TLR}
    T_1=\frac{3-4 \mu  R+3 \sqrt{1-4 \mu  R}}{\sqrt{2} \pi  R \left(1+\sqrt{1-4 \mu  R}\right)^{3/2}}\,,\qquad T_2=\frac{3-4 \mu  R-3 \sqrt{1-4 \mu  R}}{\sqrt{2} \pi  R \left(1-\sqrt{1-4 \mu  R}\right)^{3/2}}\,.
\end{equation}
Note for $\mu=0$ we have   $T_1=\frac{3}{2\pi R} $ and $T_2\rightarrow \infty$.  Further, the two temperatures are the same at the coincidence point $\mu_{coin} = 1/4R$, so that  $T_{coin}=\frac{\sqrt{2}}{\pi  R}.$\footnote{In general $d$ this coincidence point occurs at $x_{coin} = \sqrt{\frac{d-2}{d}}$, $\mu_{coin} = \frac{2}{dR} \left ( \frac{d-2}{d}\right)^{\frac{d-2}{2}}$ and $T_{coin} = \frac{\sqrt{d(d-2)}}{2 \pi R} $.}  The free energy $G$ is only real for $\mu \le \mu_{coin}$, hence there are no $G (T)$ plots for $\mu > \mu_{coin}$. 

Looking for example at the red curve in the left plot of Figure \ref{fig:GTplot}, starting from the right intersection $T_2$  the value of the parameter $x$ increases along the curve up to the cusp at $T_{0}$, and increases further as we move down from the cusp to  the left intersection $T_1$. Therefore, the CFT phase lying between the cusp and $T_2$ corresponds to a low-entropy phase and the state lying between the cusp and $T_1$ corresponds to a high-entropy phase. The temperature of the cusp can be computed by obtaining the stationary point of $T(x)$, i.e.
\begin{equation} \label{cusptemp}
\begin{aligned}
   &\left ( \frac{\partial T}{\partial x} \right)_{\mu} = 0 \quad \text{at} \quad x_0 = \left (\frac{d-2}{2}\mu R \right)^{1/d}\,,\\
   &\,\,\,\text{hence} \qquad T_{0} = \frac{d}{2\pi R} \left ( \frac{d-2}{2} \mu R\right)^{1/d}\,. 
   \end{aligned}
\end{equation}
We note that $T_0(\mu_{coin})= T_{coin}$, thus $T_0$, $T_1$ and $T_2$ are all the same for $\mu =\mu_{coin}.$

Let us now us describe the phase transitions   in this ensemble. For $\mu\le 0$ there is only a single phase, but for $0< \mu < \mu_{coin}$ a phase transition can occur between the   low- and high-entropy branches. For $T_0 < T < T_1$ the free energy is lowest for the high-entropy state, hence this phase is thermodynamically preferred in this regime.  At $T=T_1$ the high-entropy phase terminates and  there is a zeroth-order phase transition between the high-entropy and   low-entropy state.  For $T_1< T < T_2$ the low-entropy state is the only existing phase, and hence it dominates the thermodynamic ensemble. An interesting feature of the low-entropy branch is that it has positive heat capacity between  $T_2$  and some intermediate temperature
\be
T_{int}=\frac{(d-2) \mu  \left(\frac{2}{d \mu  R}\right)^{\frac{d-1}{d-2}}}{4\pi }+\frac{ (d-1) \left(\frac{2}{d \mu
    R}\right)^{-\frac{1}{d-2}}}{2\pi  R}, \label{intermediatetemp}
\ee
    which is indicated by a dashed line on the left in Figure \ref{fig:GTplot}, for which  $x_{int}=\big(\frac{d}{2}\mu R\big)^{1/(d-2)}$. But the low-entropy phase has a negative heat capacity between $T_{int}$ and $T_0$ and is hence unstable for low temperatures. The   temperatures $T_{int}$ and   $T_1$, at which the zeroth-order phase transition occurs, coincide for two values of the chemical potential, the first is $\mu_{coin}$ and the second is given by      $\mu_*=\frac{1}{324R} \left(16 \sqrt{7}-35\right)$ in $d=4$, and the associated temperature is $T_*= \frac{2 }{3 \pi R}\sqrt{2+8/\sqrt{7}}$. This means that for   $ \mu_* <\mu<\mu_{coin}$   the zeroth-order phase transition takes place between the \emph{stable} low-entropy phase and the high-entropy phase, whereas for $0<\mu <\mu_*$  the phase transition happens between the \emph{unstable} low-entropy state and  the stable high-entropy state.
    
    We would like to point out that the zeroth-order phase transition could in principle happen both ways between the high- and low-entropy state, depending on whether the temperature is decreased or increased. On the one hand, the phase transition from the high- to the low-entropy state occurs when the temperature of the CFT  increases above $T=T_1$, since above this temperature   only the low-entropy branch survives, but this transition has the peculiar feature that the entropy decreases during the process, in contradiction to the second law of thermodynamics. On the other hand, the phase transition from the low- to the high-entropy state   happens  when the temperature decreases below $T=T_1$, and this transition seems   thermodynamically more favourable since the entropy   increases during the process.

We indicated the different phases in the $\mu - T$ plane  on the right in Figure \ref{fig:GTplot}. The green region is associated  with the high-entropy phase, while the red and yellow regions correspond to the stable and unstable low-entropy phases, respectively. The zeroth-order phase transition occurs between   the high-entropy phase (green region) and the   low-entropy phase (red and yellow regions). There are no solutions in the white regions. For $\mu<0$, the (red) boundary is fixed by the value of $T_1$. For any given $\mu>0$, the left and right boundaries of the phase space are set, respectively, by the temperature of the cusp $T_0$ and the temperature $T_2$. These two boundaries meet at the coincidence point, where   $\mu_{coin} = \frac{2}{dR} \left ( \frac{d-2}{d}\right)^{\frac{d-2}{2}}$ and $T_{coin} = \frac{\sqrt{d(d-2)}}{2 \pi R} $. Further, the coexistence line (the dashed line between the green and red/yellow regions) between the high- and low-entropy phase is determined by the temperature $T_1$.  And the coexistence line between the stable and unstable low-entropy phase (the dashed line between the red and yellow regions) is set by the intermediate temperature $T_{int}$. The two coexistence lines (the intersection of the two dashed lines) meet at  the point $(\mu_*,T_*).$

\subsection{Other  ensembles}
\label{sec:otherensembles}

     \begin{figure}[t]
     \centering
     \includegraphics[scale=0.65]{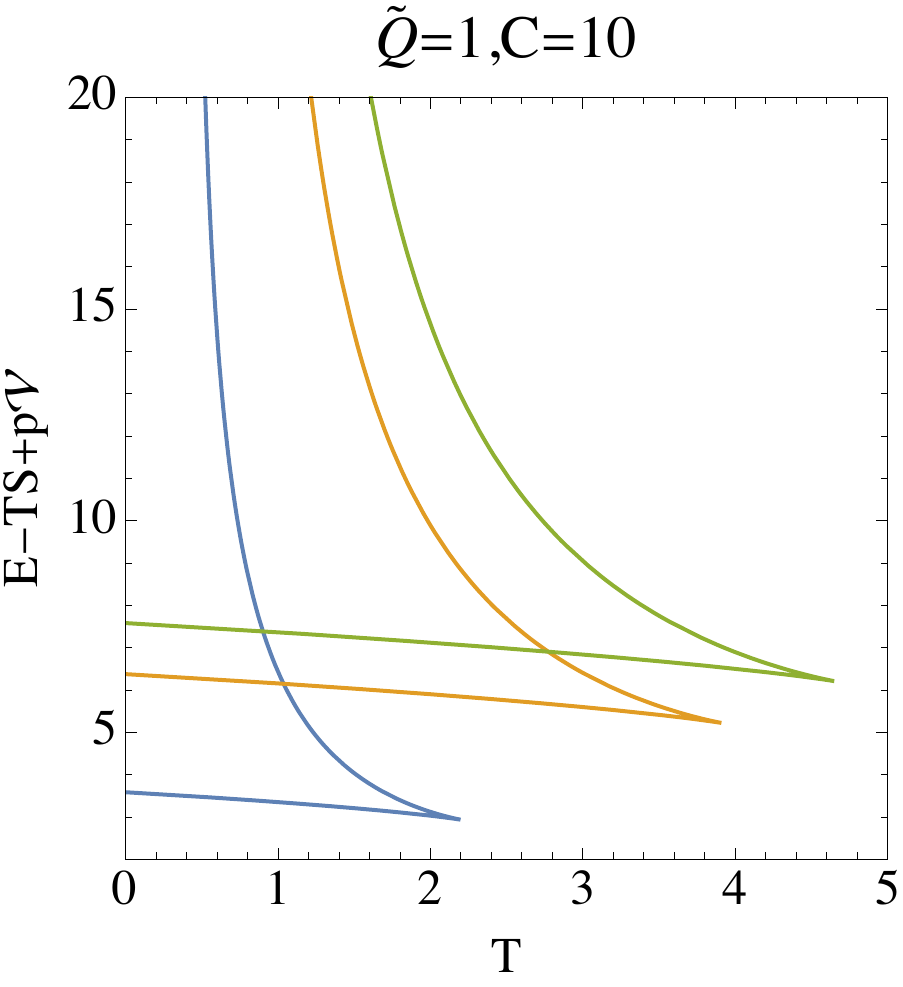}
     \includegraphics[scale=0.65]{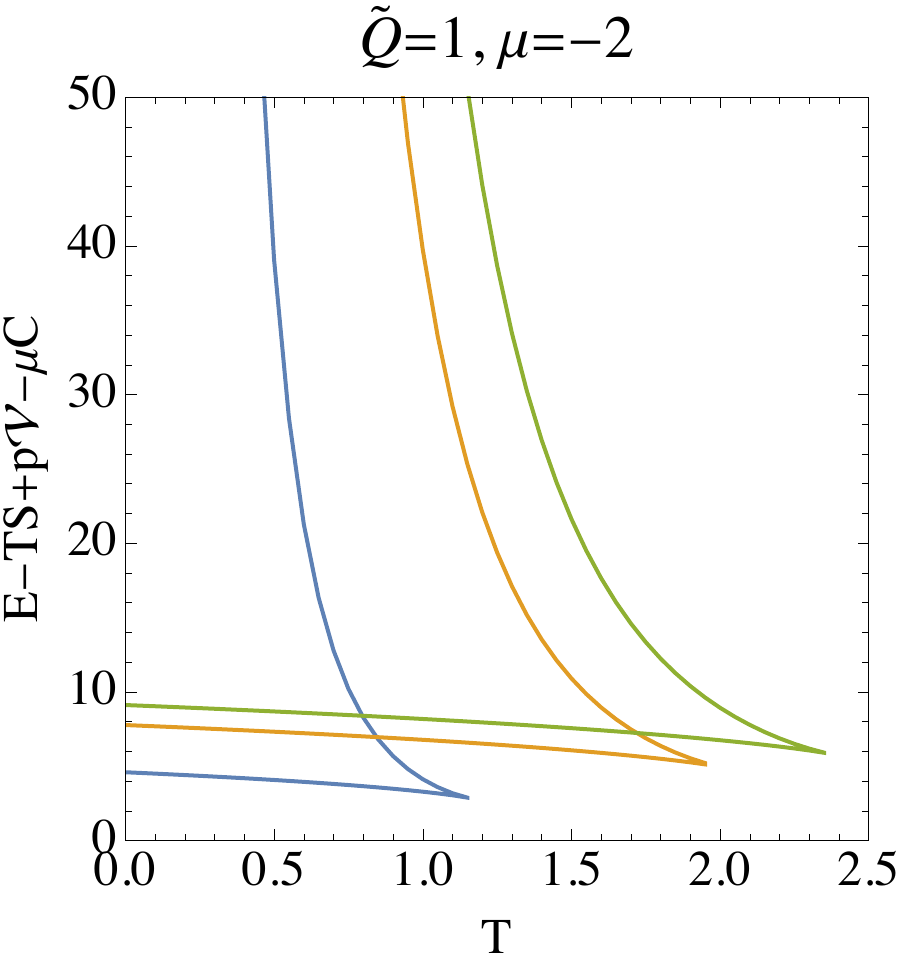}
     \includegraphics[scale=0.65]{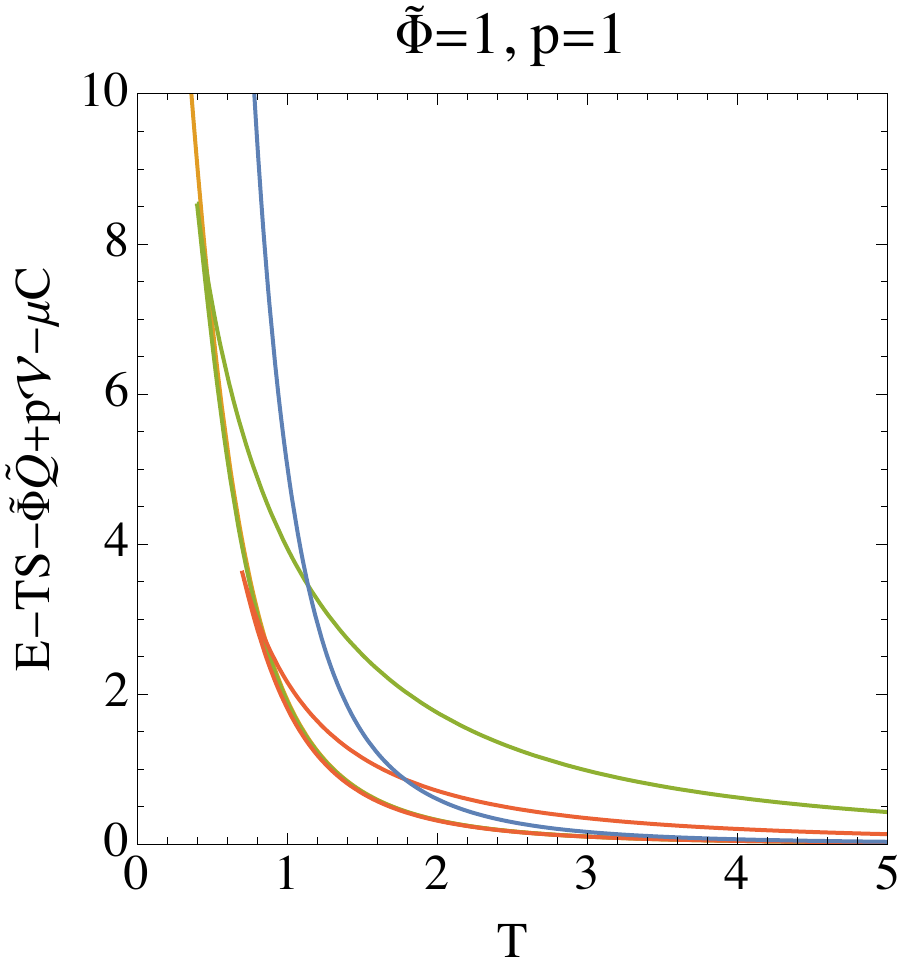}
     \includegraphics[scale=0.65]{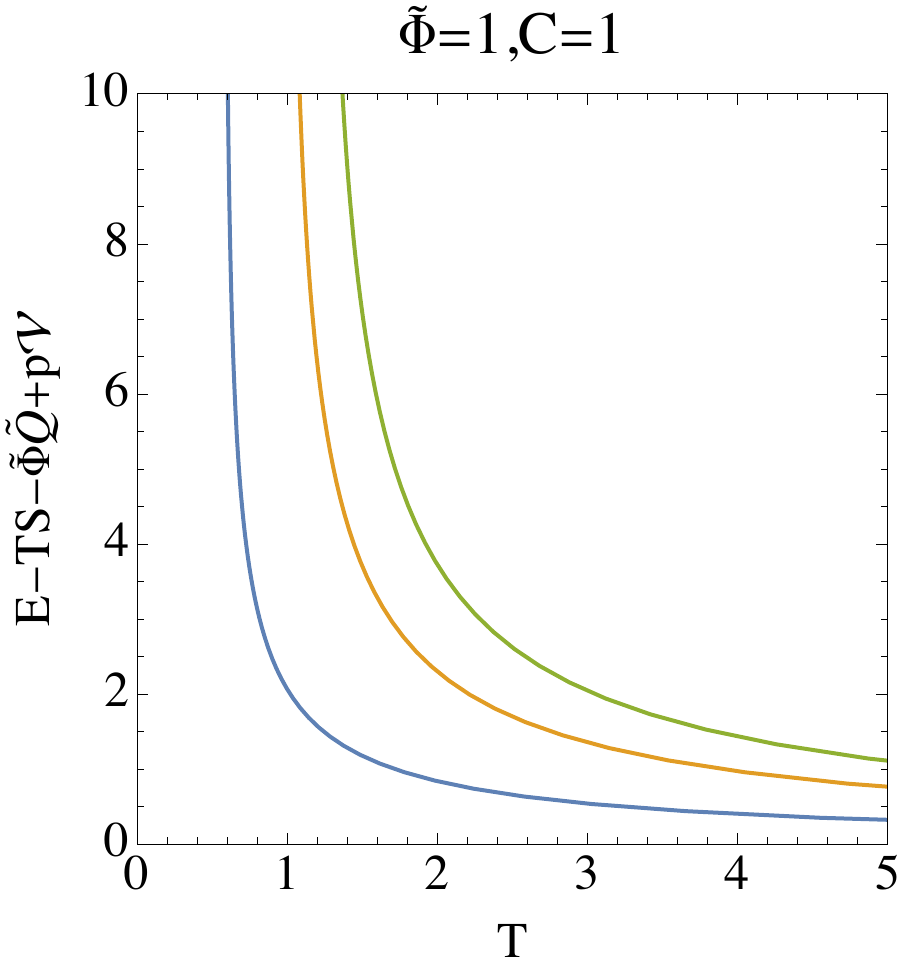}
     \caption{The   free energies of other   ensembles   plotted against temperature in $d=4$, corresponding to ensembles at fixed $( \tilde Q, p,C)$, $( \tilde Q, p,\mu)$, $(\tilde \Phi,p,\mu)$ and $(\tilde \Phi,p,C)$ (from left to right). In the top left and right diagrams   the free energies are plotted for $p=1$ (blue), $p=10$ (orange) and $p=20$ (green). On   the bottom left, the free energy   of the  fixed $(\tilde \Phi,p,\mu)$ ensemble  has qualitatively different behaviour  for $\mu\leq 0$ and $\mu>0$. The former is single branched with an asymptote at $T=0$, while the latter is double branched with a ``tip'' corresponding to a maximum $\tilde\Phi$ and minimum $T$; the parameters used are $\mu=-10$ (blue), $\mu=0$ (orange), $\mu=1/20$ (green) and $\mu=1/5$ (red). On the bottom  right,  the free energy plot for fixed $(\tilde \Phi,p,C)$ has a single branch  with an asymptote at  $T=0$; the parameters used are $p=1$ (blue), $p=5$ (orange), $p=10$ (green). These four ensembles do not display phase transitions. }
     \label{fig:others}
 \end{figure}

 In the previous subsections we discussed the (grand) canonical thermodynamic ensembles at fixed ($\tilde Q, {\cal V}, C$), ($\tilde\Phi, {\cal V}, C$) and ($\tilde Q, {\cal V}, \mu$). In addition, there are five other grand canonical ensembles, for which the corresponding free energies  $F_1 - F_5$ are given in equation \eqref{freeenergies1}.

In Figure \ref{fig:others}  the free energies are plotted against temperature   for the first four ensembles  for some representative parameter values. From left to right, these are the fixed $(\tilde Q, p,C  )$, $( \tilde Q, p,\mu)$, $(\tilde \Phi,p,\mu)$ and $(\tilde \Phi,p,C)$ ensembles.  The ensemble not displayed in the figure  is the fixed $(\tilde \Phi, {\cal V},\mu)$ ensemble, for which the associated free energy $F_5$ is   identically zero ($F_5\equiv 0$) due to the Euler equation. We further note that  in the infinite volume or infinite temperature limit  $TR \to \infty$,    two of the free energies simplify drastically  
  \begin{equation}
      T R \to \infty : \qquad F_1 = \tilde \Phi \tilde Q,  \quad F_4=0  \,,
  \end{equation}
  since we have $\mu C = - p{\cal V}$ in this limit (see Appendix   C in \cite{Visser:2021eqk}).
    Moreover, in this limit the free energy $F_1$ coincides with the free energy $G$ \eqref{eq:freeenergy3} in the fixed $(\tilde Q, {\cal V}, \mu)$ ensemble, i.e. $F_1 =  \tilde \Phi \tilde Q=G$. And, in this limit the free energy $F_2$ is equal to minus the free energy $W$ \eqref{eq:freeenergy2} in the fixed $(\tilde \Phi, {\cal V}, C)$ ensemble, i.e. $F_2 = - \mu C = - W$. 
  
   Importantly, from Figure \ref{fig:others}  we see that all  five ensembles do not display critical phenomena or phase transitions. Therefore, there is no critical behaviour, such as $p - \cal V$ criticality, in the fixed $p$ ensembles in the CFT. Further, in the first three ensembles at fixed $(\tilde Q, p,C  )$, $( \tilde Q, p,\mu)$ and $(\tilde \Phi,p,\mu)$, there are two branches in the free energy plots: the lower branch corresponds to a thermal state with a high value of $S/C$, dual to a large black hole in the bulk, and the upper branch is associated with a thermal state with a low value of $S/C$, coresponding to a small AdS black hole.  The lower branch in these plots has the lowest free energy for a fixed temperature, and hence  this phase  dominates the respective ensembles;  moreover, it has   positive specific heat and is therefore stable against thermal fluctuations. The upper branch     has   negative specific heat, and is therefore thermodynamically unstable.

  \section{CFT criticality and thermodynamic stability}
\label{sec:critistabl}

In this section we study the critical behaviour of the (extended) CFT thermodynamics corresponding to charged AdS black holes. We show that the critical points in       the $\tilde Q - \tilde \Phi$ and $C- \mu$ planes have mean field critical exponents. Further, we derive the heat capacities for the ensembles at fixed $(\tilde Q, {\cal V}, C)$, $(\tilde \Phi, {\cal V}, C)$ and $(\tilde Q, {\cal V}, \mu)$, and analyse the thermodynamic stability of the different phases in these ensembles.

\subsection{Critical points and comparison with Van der Waals fluid}
   
 As is widely known, the critical behaviour of charged AdS black holes resembles that of common thermodynamic fluids. From what we have just seen, this resemblence is maintained in the dual CFT. In the fixed $(\tilde Q, {\cal V}, C)$ ensemble the first-order phase transition between low- and high-entropy states observed above is similar to the first-order liquid-gas phase transition of Van der Waals fluids. If one plots the three-dimensional $\tilde G-T-P$ diagram for the Van der Waals fluid (see for example Figure 3 in  \cite{Kubiznak:2012wp}), where $\tilde G\equiv E-TS+PV$ is the Gibbs free energy, one obtains a swallowtail diagram similar to what we would obtain if plot the $F-T-\tilde Q$ or $F-T-C^{-1}$ diagram. Therefore, we can view respectively $\tilde Q$ and $1/C$   as playing a role  analogous to the pressure of the Van der Waals fluid in driving the system to its critical point. Moreover,  the electric potential $\tilde \Phi$ and the chemical potential $\mu$ are analogous to  the volume of the Van der Waals system.  In fact, critical exponents of the CFT critical point are the same as the critical exponents of the Van der Waals fluid, and hence they fall within the same   universality class. We will perform this computation in both the $\tilde Q-\tilde \Phi$ and $C-\mu$ plane in the next subsection.
 
  \begin{figure}
    \centering
    \includegraphics[scale=0.75]{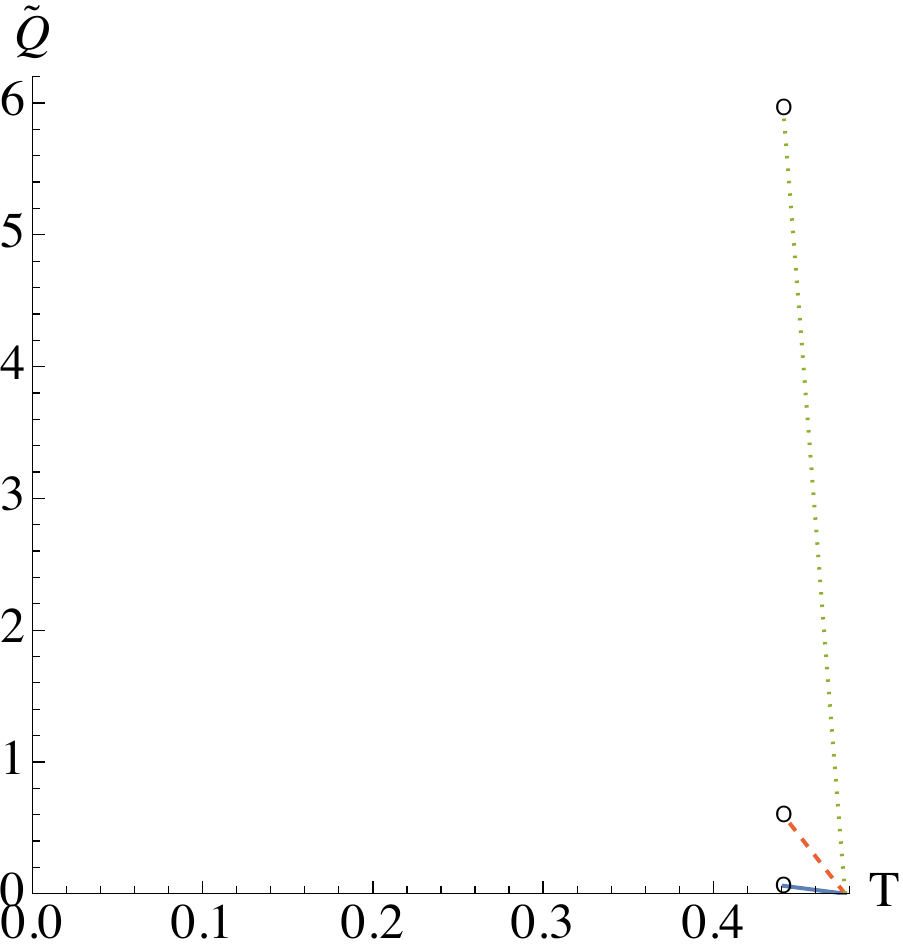} \hspace{1 cm}
    \includegraphics[scale=0.75]{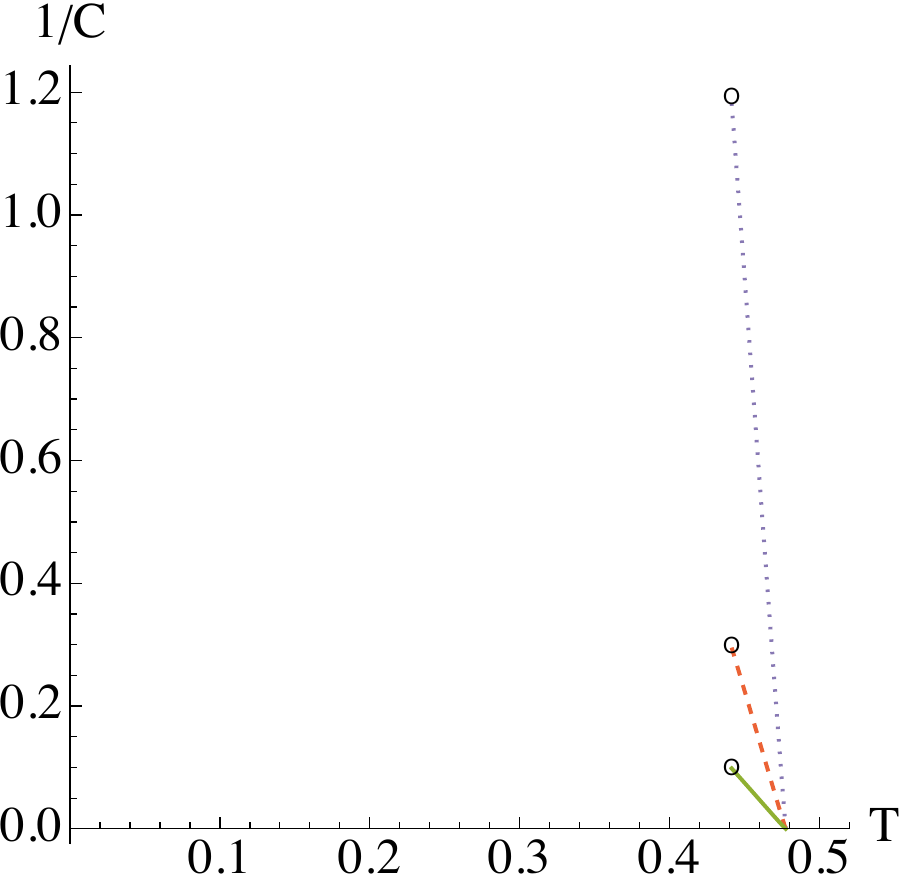}
    \caption{Coexistence lines in $d=4$ for the fixed $(\tilde Q,{\cal V},C)$ ensemble.  \textbf{Left}: Low-entropy and high-entropy   coexistence curve for CFT thermal states on $\tilde Q-T$ phase diagram. The parameters used here are $R=1$, $C=1/10$ (blue), $C=1$ (red, dashed) and $C=10$ (green, dotted). For each value of~$C$, the coexistence line represents a line of first-order phase transitions between low-entropy states (to the left of the line) and  high-entropy states (to the right), and the line ends at a critical point where a second-order phase transition occurs at $\tilde Q=\tilde Q_{crit}$ and $T=T_{crit}$.  \textbf{Right}: A similar coexistence curve of the low- and high-entropy CFT states exists in the $1/C-T$ phase diagram, which we depicted here for  $R =1$ and  $\tilde Q=1/2$ (green), $\tilde Q=2$ (orange, dashed) and $\tilde Q=6$ (blue, dotted).} 
    \label{fig:qvcFT2}
\end{figure}

 Let us first compute the values of the relevant thermodynamic quantities at the critical point. The temperature has an inflection point as a function of $x$  at the critical point \cite{Chamblin:1999tk}, i.e.
 \begin{equation}
 	\left (\frac{ \partial T}{\partial x} \right)_y= 0  =	\left ( \frac{\partial^2 T }{\partial x^2} \right)_y\, \qquad \text{at} \quad  x=x_{crit}  \quad \text{and} \quad y=y_{crit}.
 \end{equation}
 From \eqref{energytempcft}  we find that these relations are solved by 
 \begin{equation}
 \label{critpt}
 	x_{crit}  =	\frac{(d-2) }{\sqrt{d(d-1)}} \,,\qquad   y_{crit}  = \frac{1}{\sqrt{(d-1)(2d-3)}} x_{crit}^{ d-2 }\,.
 \end{equation}
This yields a critical value for the ratio of the central charge and the electric charge
 \begin{equation}
 \label{Ccrit}
 	\frac{C_{crit}}{\tilde Q_{crit}} =\frac{1}{2 \alpha (d-1) y_{crit}  }   = \sqrt{\frac{ (2d-3)}{8(d-2)  }}\frac{1}{x_{crit}^{d-2}} \, .
 \end{equation}
 For $d=3$ we have $C_{crit} /\tilde Q_{crit}= 3/2$ and for $d=4$ we have $C_{crit}/\tilde Q_{crit} =  3 \sqrt{5} / 4 . $   
 When the electric  charge vanishes $\tilde Q=0$ there is no critical point and hence no critical value of the central charge.

The entropy \eqref{SQphi} and  temperature \eqref{energytempcft} at the critical point are
 \begin{equation}
 \label{Tcrit}
 S_{crit}= 4\pi  \tilde Q_{crit}\sqrt{\frac{ (d-2)(2d-3)}{8d(d-1)  }}\, ,  \quad	T_{crit} = \frac{(d-2)^2}{(2d-3)\pi R x_{crit}} =\frac{d-2}{2d-3} \frac{\sqrt{d(d-1)}}{\pi R}\,.
 \end{equation}
 For completeness, we also write down the critical values of 
 the chemical potential \eqref{mu}, the electric potential \eqref{eq:electricpotential1} and the pressure \eqref{pmu}
 \begin{align}
 	\mu_{crit} &= \frac{6 (d-2)}{d (2d-3)} \frac{x_{crit}^{d-2}}{R} \,, \qquad
 \tilde \Phi_{crit}	= \frac{1}{ R \sqrt{2(d-2)(2d-3)}}\,,  \\ 
 	p_{crit}  &= \frac{C_{crit}x_{crit}^{d-2} }{\Omega_{d-1}R^d} \frac{4(d (d-3)+3)}{d(2d-3)} = \frac{\tilde Q_{crit}}{\Omega_{d-1}R^d}   \frac{2(d (d-3)+3)}{d\sqrt{ 2(d-2)(2d-3)   }  } \,.
 \end{align}
 Figure \ref{fig:qvcFT2} shows the coexistence lines of the low- and high-entropy phases of the CFT on the $\tilde Q-T$ as well as the $1/C-T$ phase diagrams. The coexistence line separates the two phases on these planes and the CFT undergoes a first-order phase transition as it crosses the line. On both of these phase diagrams, the low-entropy phase lies to the left of the coexistence line (for any given value of $C$ and $R$ respectively) while the high-entropy phase lies to the right. The critical points are depicted by open circles on the diagram. Above the critical points, the CFT does not display distinct phases. 
 
 One point to note about   
 Figure \ref{fig:qvcFT2} is that while   $\tilde Q$ and $1/C$ in the CFT can be roughly viewed as playing the analogous role of the pressure of a Van der Waals fluid, we see a slight qualitative difference  in the respective coexistence lines.   Those in the CFT are  negatively  sloped, whereas those   in   the $P-T$ plane of a Van der Waals fluid (and the $1/C-T$ of bulk charged AdS black holes in \cite{Cong:2021fnf}) are positively  sloped. This mismatch in the analogy can be resolved by replacing the temperature by the inverse temperature  \cite{Chamblin:1999tk,Chamblin:1999hg,Kubiznak:2012wp}. That is, the coexistence lines in the $\tilde Q - \beta$ and $1/C - \beta$ phase diagrams have a positive slope, and are therefore more analogous to the those in the $P-T$ phase diagram of a Van der Waals fluid (see for instance Figure 13 in \cite{Kubiznak:2012wp} for the coexistence curve in  the $ Q - \beta$ plane).  In Table \ref{tableanalogies} we summarize  the two formal analogies between the CFT thermal states and the Van der Waals fluid. This shows the   analogies are purely mathematical and do  not identify the same physical quantities. Therefore, we conclude   CFT states dual to charged AdS black holes are not identical to Van der Waals fluids, they just fall within the same universality class. 
 
     \begin{table}[t!!]
        \centering
\begin{tabular}{|c||c|c|c |c|   }
 \hline
Van der Waals fluid& analogy 1 & analogy 2  \\
 \hline
 temperature   & $\beta$   &   $\beta$ \\
 pressure  &   $\tilde Q$  & $1/C$   \\
 volume  &   $\tilde \Phi$   & $\mu$   \\
 \hline
\end{tabular}
         \caption{Two analogies between the thermodynamic variables of the Van der Waals fluid and those of the CFT thermal state dual to charged AdS black holes.
         }
    \label{tableanalogies}
   \end{table}
 
\begin{figure}
    \centering
    \includegraphics[scale=0.75]{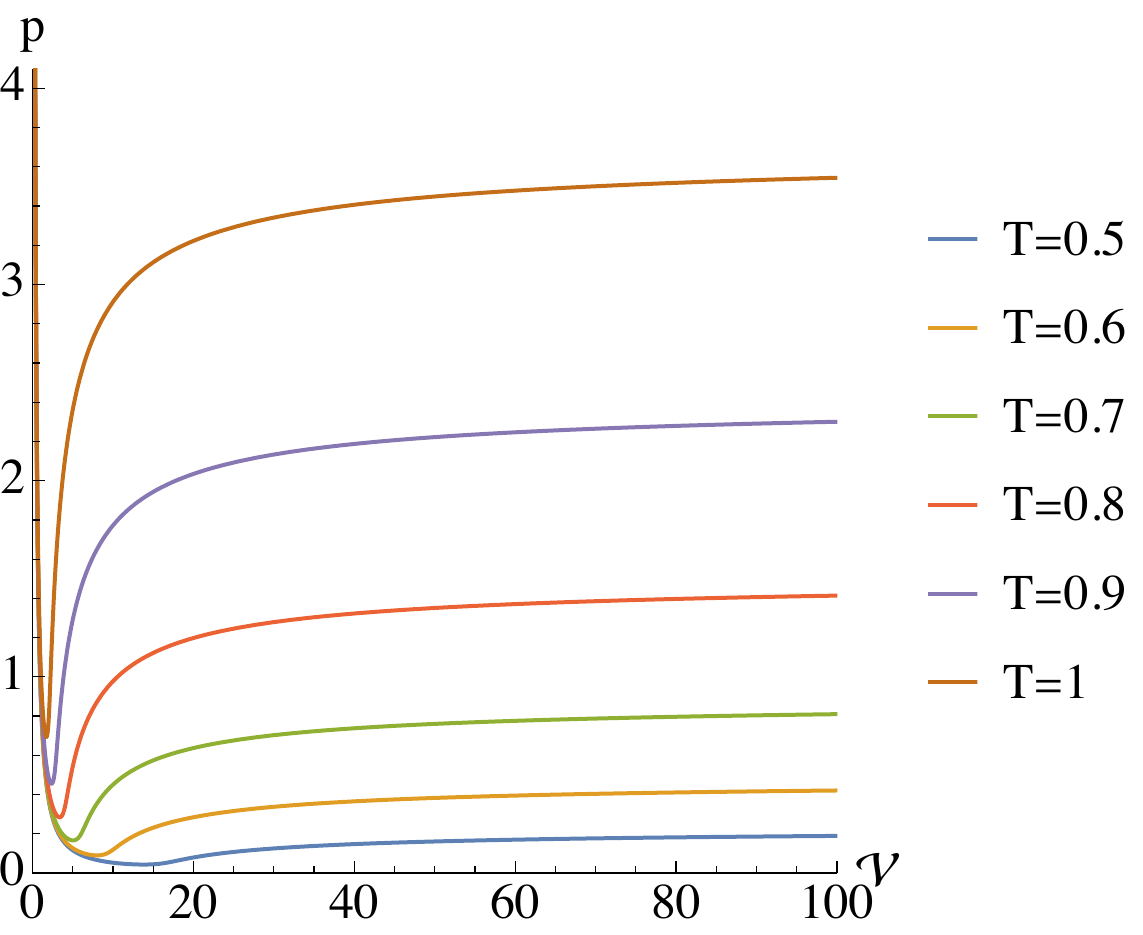} 
    \caption{No $p-{\cal V}$ criticality for   holographic CFTs. The $p-\cal{V}$ phase diagram for the CFT is very different from that of the Van der Waals fluid. At fixed $C$ and $\tilde Q$, the diagram shows the same qualitative behaviour for all $T$. In particular, there is no critical temperature at which the curve displays an inflection point. The parameters used here are $\tilde Q =C= 1$ and $d=4$, but the same qualitative behaviour applies  in other dimensions   $d>2$.}
    \label{fig: pressure plots}
\end{figure}

 Finally, since  the critical point given by  \eqref{critpt} does not depend on $\cal V$, and no critical points were found in the fixed $p$ ensembles, there is no critical behaviour in the $p-{\cal V}$ plane in the CFT. This stands in stark contrast to the     $P-V$ criticality found in  \cite{Kubiznak:2012wp} for charged black holes using the standard interpretation of extended black hole thermodynamics, where $P$ is the bulk pressure \eqref{P} and $V$ the associated thermodynamic volume. For illustration, we display the CFT $p({\cal V})$   diagram in Figure \ref{fig: pressure plots} for different values of the temperature at fixed $(\tilde Q, C)$. We see that the behaviour is different from that of the Van der Waals fluid (see for instance Figure 1 in \cite{Kubiznak:2012wp}). For any temperature the pressure $p$ first decreases to a global minimum before increasing with the volume $\cal{V}$ for the given $(\tilde Q, C)$. In particular, there is no critical temperature at which the $p(\cal{V})$ plot displays an inflection point. In other words, the CFT fluid dual to a charged black holes is \emph{not} a standard Van der Waals fluid.

 \subsection{Critical exponents for fixed $(\tilde Q, {\cal V}, C)$ ensemble}
 \label{sec:criticalexp}
 In the previous subsection  we have seen that there is no criticality in the $p-\cal{V}$ plane; the   critical point only appears in the $C-\mu$ and $\tilde Q-\tilde\Phi$ plane. Here, we aim to derive the   critical exponents of the critical points in these respective planes.

\subsubsection{Criticality in $C-\mu$ plane}

 First we study the   criticality in the $C-\mu$ plane.
  We can   express the central charge   and chemical potential  as the functions $C=C(\tilde Q, T, R, x)$ and $\mu = \mu(T,R,x)$, respectively, 
 \begin{align}
 \label{muc}
 	C&= \frac{\tilde Q }{2 \alpha (d-1) x^{ d-2}\sqrt{  1+\frac{d}{d-2} x^{ 2}-T x \frac{4 \pi R}{d-2}  }}   \,, \\
 	\mu &= \frac{x^{d-1}}{R} \left ( T   \frac{4 \pi R}{d-2}  - \frac{2(d-1)}{d-2} x     \right) \,.
 \end{align}
 The ``equation of state'' is described by $C=C(  \mu,T)$ for fixed $(\tilde Q, V)$, since $C$ (or actually $1/C$) is analogous to pressure and $
 \mu$ to volume in the second analogy with the Van der Waals fluid in Table \ref{tableanalogies}. In Figure \ref{fig: cmuplot} we plot $C(\mu)$ parametrically using $x$ as parameter for different values of $T$. For each temperature, there exists a maximum value $\mu_{max}$ for the chemical potential. This happens at $x  = \frac{4 \pi }{2d}T R   $ where we have $\mu_{max} = \frac{2}{(d-2)R} x^d$. The local features of the critical point are more easily visible in the right diagram of Figure \ref{fig: cmuplot}. The critical point is indicated here by the black dot, which is the inflection point of the $C(  \mu,T_{crit})$ curve. Above the critical temperature  the $C(\mu)$ curve displays a  ``wriggle'', i.e. a local minimum is followed by a local maximum. To describe the first-order phase transition between high- and low-entropy CFT phases in the fixed $(\tilde Q, {\cal V}, C)$ ensemble one has to  replace  the wriggle by a horizontal $C=\text{constant}$   line, in accordance with  Maxwell's equal area law \cite{Kubiznak:2012wp}. This means that the areas   above and below the  $C=\text{constant}$  line  (marked in red in Figure \ref{fig: cmuplot}) are equal. The first-order phase transition   takes place across this  $C=\text{constant}$ line. 
   
To study the critical exponents in the $C-\mu$ plane, let us define the variables
\be
t \equiv \frac{T-T_{crit}}{T_{crit}}\,,\qquad
\chi \equiv \frac{C}{C_{crit}}\,,
\qquad
\psi \equiv \frac{\mu-\mu_{crit}}{\mu_{crit}} \,.
\ee
Note that we are not precisely following the second analogy with the Van der Waals fluid in Table~\ref{tableanalogies}, since we are using $T$ and $C$ as our variables instead of $\beta$ and $1/C$ respectively, but this is not necessary to derive the mean field critical exponents.  The critical exponents   that bear our interest are defined as follows:

  \begin{figure}
     \centering
     \includegraphics[scale=0.75]{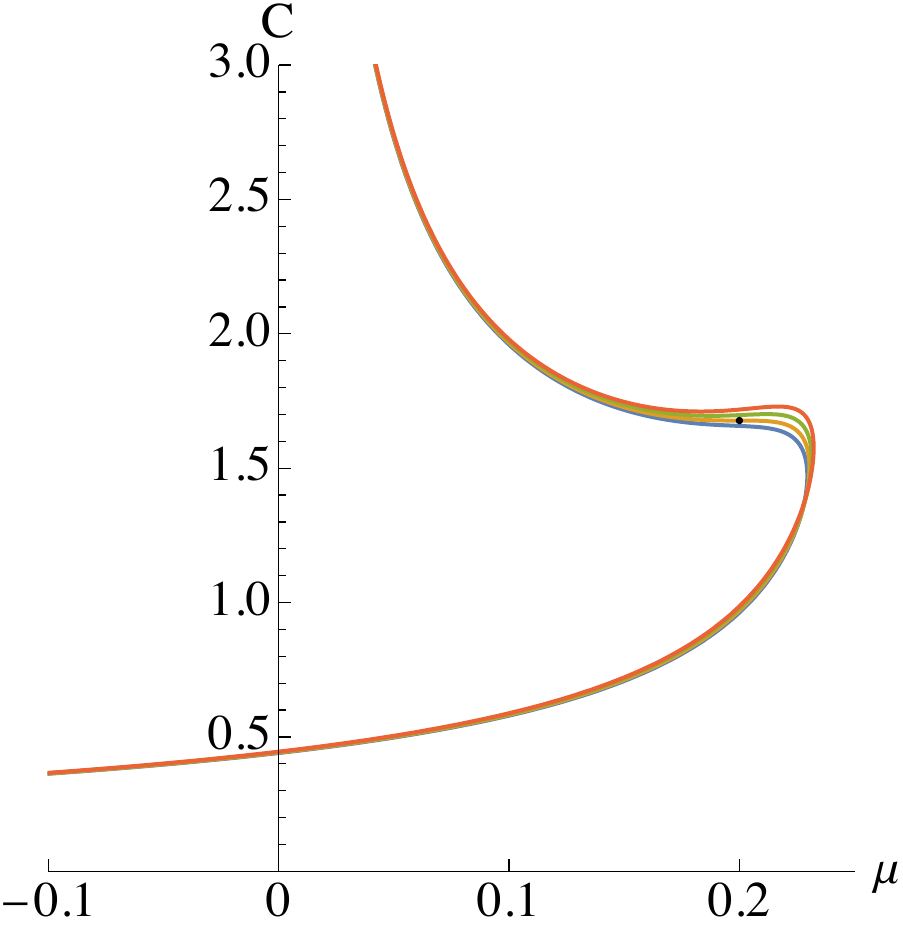}
     \includegraphics[scale=0.75]{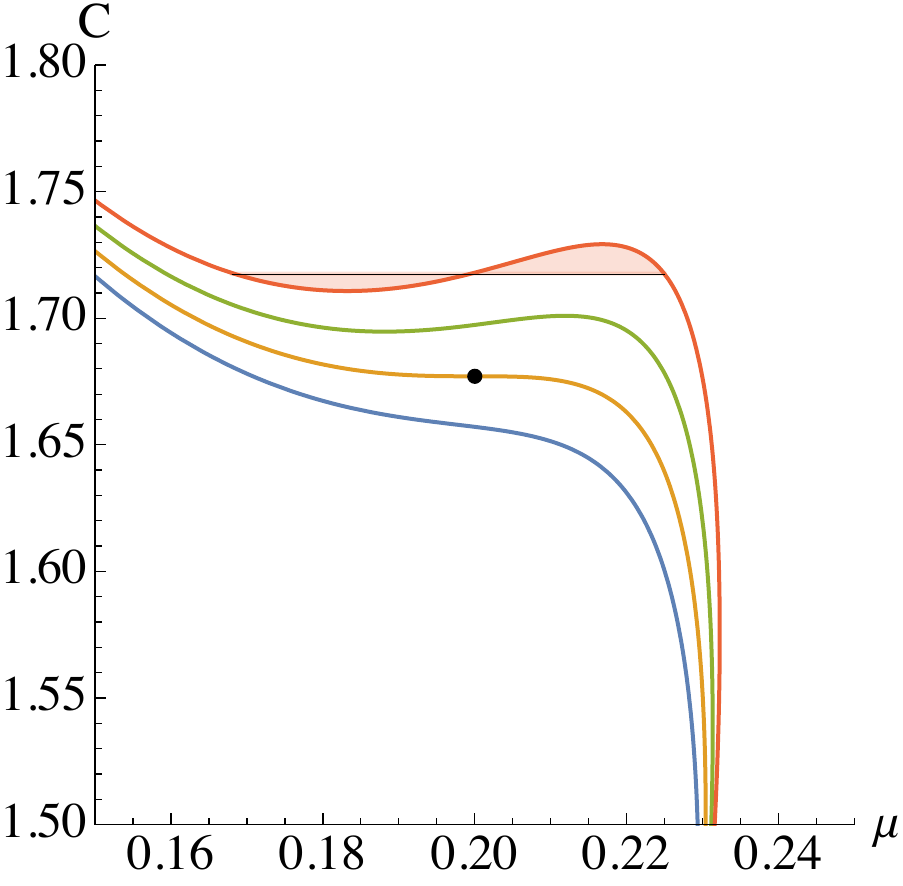}
     \caption{Isotherms in the $C-\mu$ plane for $d=4$. The parameters used here are $\tilde Q=1$, $R=1$ and  $T=T_{crit}=\frac{4\sqrt{3}}{5\pi}$ (orange), $T=0.999T_{crit}$ (blue), $T=1.001T_{crit}$ (green) and $T=1.002T_{crit}$ (red). The right plot shows a zoomed-in version of the left plot around the critical point (black dot). For $T>T_{crit}$ (e.g. red curve), the curve displays a ``wriggle'' similar to the familiar $P-V$ phase diagram in the Van der Waals liquid-gas phase transition. The phase transition takes place in accordance with Maxwell's equal area law.}
     \label{fig: cmuplot}
 \end{figure}

\begin{itemize}
    \item The critical exponent $\alpha$ governs the behaviour of the specific heat at constant $(\tilde Q, {\cal V}, \mu)$,
    \be {\cal C}_{\tilde Q, {\cal V}, \mu} \equiv T \left ( \frac{\partial S }{\partial T} \right)_{\tilde Q, {\cal V}, \mu}\sim |t|^{-\alpha}\,.
\ee
\item The exponent $\beta$ describes the behaviour of the order parameter $\eta \equiv \mu_h-\mu_l$ where $\mu_h$ ($\mu_l$) is the value of $\mu$ of the high-entropy (low) CFT state at the first-order transition,
\be
\eta \equiv \mu_h-\mu_l\sim |t|^{\beta}\,.
\ee
\item The exponent $\gamma$ determines the behaviour of $\kappa_T$, 
\be
\kappa_T \equiv -\frac{1}{\mu} \left ( \frac{\partial \mu }{\partial C} \right)_{T,{\tilde Q, {\cal V}}} \sim |t|^{-\gamma}\,,
\ee
where we fix the temperature but do not
hold $\mu$ fixed.
\item The exponent $\delta$ governs the following behaviour on the critical isotherm $T=T_{crit}$  
\be
|C-C_{crit}| \sim |\mu-\mu_{crit}|^{\delta}\,.
\ee
\end{itemize}
First, to find $\alpha$ we can use the following expressions for the entropy  $S(\tilde Q, R, \mu,x)$ and the temperature $T(\tilde Q, R, \mu,x)$,
\begin{align}
    S &= \frac{2 \pi  \tilde Q x^{\frac{d}{2}+1}}{\sqrt{(d-2)(d-1)}  \sqrt{-2 \left(x^2-1\right) x^d-2 \mu  R x^2}}\,, \\ 
    T&=\frac{x }{4 \pi  R}\left((d-2) \mu  R x^{-d}+2 (d-1)\right)\,,
\end{align}
obtained by substituting $C$ and $y$ in terms of $\mu$,   $\tilde Q$ and $x$, using  \eqref{SQphi} and  \eqref{mu}. This gives,
\begin{equation}
\begin{aligned}
    {\cal C}_{\tilde Q,{\cal V},\mu} &= \frac{\pi  \tilde Q x^{\frac{d}{2}+1} \left(2 x^d-d \mu  R x^2\right) \left((d-2) \mu  R+2 (d-1) x^d\right)}{\sqrt{2 (d-2)} (d-1)^{3/2} \left(2
   x^d-(d-2) \mu  R\right) \left(-x^{d+2}+x^d-\mu  R x^2\right)^{3/2}}
  \\ &\xrightarrow[\substack{x\to x_{crit} \\ \mu\to \mu_{crit}}]{} 2 \pi  (d-3) (d-2) (d-1) \sqrt{\frac{4 d-6}{d \left(d^2-3 d+2\right)}} \tilde Q \quad \implies \quad \alpha = 0\,.
\end{aligned}
\end{equation}
In other words, the heat capacity at constant $(\tilde Q, {\cal V},\mu)$ has a finite limit at the critical point and hence $\alpha=0$.

For the other critical exponents (see e.g. \cite{Kubiznak:2012wp}), we write \eqref{muc} in terms of $\chi =  \chi(t, \psi)$, and expand around the critical point, $t = \psi = 0$. This gives us 
\begin{equation}
\label{eos2}
\chi = 1+2 (d-2)(d-1)\, t+\frac{6(d-2)(d-1)(2d-3)}{d} \,t\,\psi-\frac{54(d-2)(d-1)(2d-3)}{d^3}\psi^3 + O(t\psi^2,\psi^4). 
\end{equation}
Differentiating this at fixed $t$ gives
\be
dC = C_{crit}\bigg(\frac{6(d-2)(d-1)(2d-3)}{d} \,t-\frac{162(d-2)(d-1)(2d-3)}{d^3}\psi^2\bigg)d\psi\,.
\ee
Let us now denote $\psi_h = (\mu_h-\mu_{crit})/\mu_{crit}$ and $\psi_l = (\mu_l-\mu_{crit})/\mu_{crit}$. Using Maxwell's equal area law and the fact that during the phase transition $\chi $ remains constant, we have the following two equations:
\begin{align}
\chi &= 1+2 (d-2)(d-1)\, t+\frac{6(d-2)(d-1)(2d-3)}{d} \,t\,\psi_h-\frac{54(d-2)(d-1)(2d-3)}{d^3}\psi_h^3 \nonumber \\
&= 1+2 (d-2)(d-1)\, t+\frac{6(d-2)(d-1)(2d-3)}{d} \,t\,\psi_l-\frac{54(d-2)(d-1)(2d-3)}{d^3}\psi_l^3\,,  \nonumber\\
0 &= \int_{\psi_l}^{\psi_h} \psi \bigg(\frac{6(d-2)(d-1)(2d-3)}{d} \,t-\frac{162(d-2)(d-1)(2d-3)}{d^3}\psi^2\bigg)d\psi\,,
\label{betaE}
\end{align}
which admit the unique non-trivial solution, $\psi_h=-\psi_l =  \frac{d\sqrt{t}}{3}$. Hence we have
\be
\eta = \mu_{crit}(\psi_h-\psi_l) = \mu_{crit} \frac{2d\sqrt{t}}{3}\quad\implies\quad \beta = 1/2\,.
\ee
To calculate $\gamma$, we differentiate \eqref{eos2}, which gives
\be
\left ( \frac{\partial \mu }{\partial C} \right)_{T, \tilde Q, {\cal V}}= \frac{ \mu_{crit}}{y_{crit}}\frac{d}{6(d-2)(d-1)(2d-3)}\frac{1}{t}+O(\psi)\,.
\ee
Hence,
\be
\kappa_T \sim \frac{d}{6(d-2)(d-1)(2d-3)y_{crit}}\frac{1}{t}\quad \implies \quad \gamma = 1\,.
\ee
Finally, we can determine $\delta$ by setting $t=0$ in \eqref{eos2}, which gives us
\be
\chi= 1+\frac{54(d-2)(d-1)(2d-3)}{d^3}\psi^3 \quad \implies \quad \delta = 3\,.
\ee
The critical exponents obtained here agree with those found previously in \cite{Kubiznak:2012wp} as well as the predictions of mean field theory (MFT).

\subsubsection{Criticality in $\tilde Q-\tilde \Phi$ plane}

\begin{figure}
    \centering
    \includegraphics[scale=0.75]{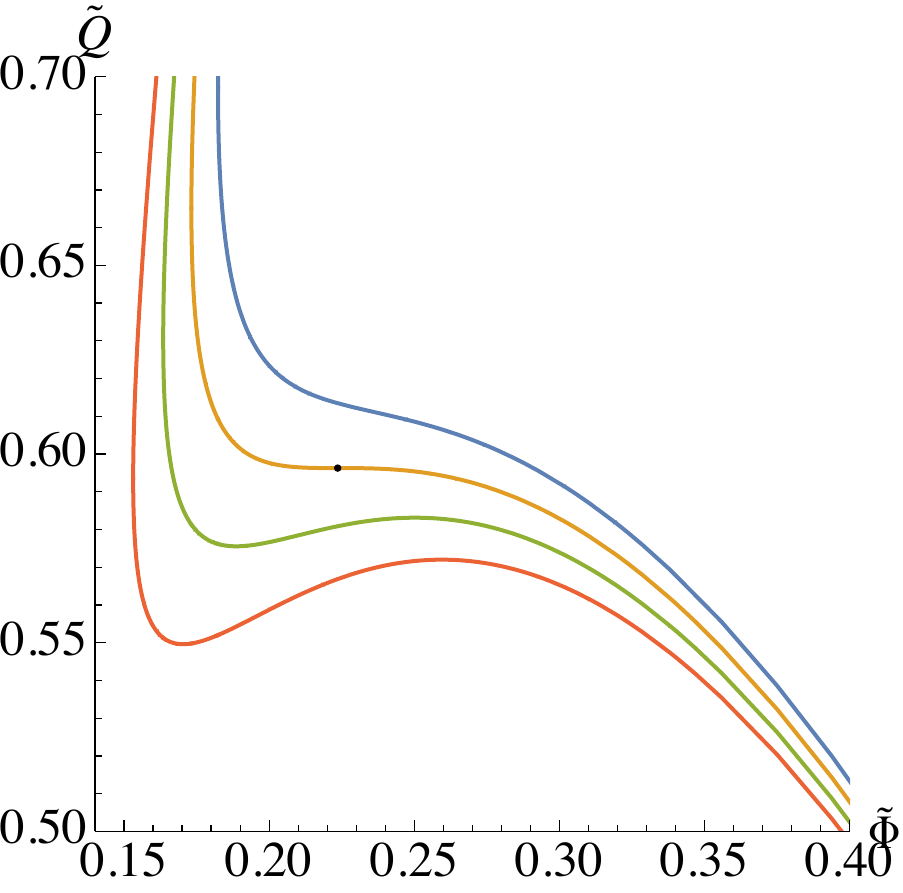}
    \caption{Isotherms in the $\tilde Q-\tilde \Phi$ plane around the critical point (black dot) for $d=4$.   The parameters used here are $C=R=1$ and $T=0.999\, T_{crit}$ (blue), $T=T_{crit}=\frac{4\sqrt{3}}{5\pi}$ (orange), $T=1.001\, T_{crit}$ (green) and $T=1.002\, T_{crit}$ (red). For the temperatures $T< T_{crit}$,  the  $\tilde Q(\tilde \Phi)$ curve displays a ``wriggle'' for which Maxwell's equal area law holds.}
    \label{fig:Qphi}
\end{figure}
We can also study the   criticality in the $\tilde Q-\tilde \Phi$ plane (see also \cite{Niu:2011tb,Dolan:2016jjc} for a similar analysis). In this case, we consider the ``equation of state''  $\tilde Q = \tilde Q(\tilde\Phi,T)$ for fixed $({\cal V},C)$, following the first analogy with the Van der Waals fluid in Table \ref{tableanalogies}. As before, we do this parametrically, using $x$ as the parameter. This gives,
\begin{align}
    \tilde Q &= 2 \sqrt{2} C x^{d-2} \sqrt{(d-1) \left(d x^2+d-2-4 \pi  R T x\right)} \,, \\
    \tilde \Phi &= \frac{    \sqrt{(d-1)(d x^2+d-2-4 \pi  R T x)}}{\sqrt{2  } (d-2) R  }\,.
\end{align}
The $\tilde Q-\tilde \Phi$ plot around the critical point is shown in Figure \ref{fig:Qphi}. Similar to the $C-\mu$ case, the curve displays a ``wriggle''  near the critical point for temperatures $T<T_{crit}$, where an equal area law holds. 

To study the critical exponents we denote
\begin{equation}
  t \equiv \frac{T-T_{crit}}{T_{crit}}, \qquad  \tilde q \equiv \frac{\tilde Q}{\tilde Q_{crit}}\,,\qquad \phi \equiv \frac{\tilde \Phi-\tilde \Phi_{crit}}{\Phi_{crit}}\,.
\end{equation}
The four critical exponents $\alpha$, $\beta$, $\gamma$ and $\delta$  in this case are defined as
\begin{align}
    {\cal C}_{\tilde \Phi, {\cal V}, C} &\equiv T \left ( \frac{\partial S }{\partial T} \right)_{\tilde \Phi, {\cal V}, C}\sim |t|^{-\alpha}\,, \qquad \quad \eta \equiv \tilde \Phi_h-\tilde \Phi_l\sim |t|^{\beta}\, ,\\
    \kappa_T &\equiv  -\frac{1}{\tilde \Phi} \left ( \frac{\partial \tilde \Phi }{\partial \tilde Q} \right)_{T, {\cal V}, C} \sim |t|^{-\gamma}\,, \qquad |\tilde Q-\tilde Q_{crit}| \sim  |\tilde \Phi-\tilde \Phi_{crit}|^{\delta}\,.
\end{align}
Compared to   the Van der Waals fluid,  here $ {\cal C}_{\tilde \Phi, {\cal V}, C} $ is the analog of the   heat capacity at constant volume, $\eta$ is the order parameter on an isotherm describing the difference  $\tilde \Phi_h - \tilde \Phi_l$ between the ``volume'' of high- and low-entropy phases, $\kappa_T$ is the equivalent of isothermal compressibility, and the exponent $\delta$ is a property of the critical isotherm  $t=0$.

The first critical exponent $\alpha$ can be derived by evaluating the heat capacity at the critical point, which yields a finite value 
\begin{equation}
\begin{aligned}
     {\cal C}_{\tilde \Phi, {\cal V}, C} &=\frac{4 \pi  C (d-1) x^{d-1} \left((d-1) \left(d x^2+d-2\right)-2 (d-2)^2 R^2 \tilde\Phi^2\right)}{d^2 (x^2-1)+2 (d-2)^2 R^2 \tilde\Phi^2-d x^2+3 d-2}
     \\ &\xrightarrow[\substack{x\to x_{crit} \\  }]{} \frac{4 \pi  C (d-1) x_{crit}^{d-1} \left(2 (d-2) \tilde \Phi^2 R^2-2
   d+3\right)}{1-2 (d-2) \tilde \Phi^2 R^2}\quad \implies \quad \alpha = 0\,.
\end{aligned}
\end{equation}
Further, in order to compute $\beta, \gamma$ and $\delta$, we need to expand   $\tilde q(t,\phi)$ near the critical point 
\begin{equation}
    \tilde q = 1-2(d-2)(d-1)\, t+2(d-1)(2d-3)\,t \phi -\frac{2 (d-1) (2 d-3)}{(d-2)^2} \phi^3\,.
\end{equation}
Following the same steps as above in the $C-\mu$ plane, we can deduce from this expansion that once again the critical exponents agree with those of MFT: $\alpha=0, \beta = 1/2, \gamma=1, \delta =3$.

  \subsection{Heat capacities and thermal stability}
\begin{figure}
    \centering
    \includegraphics[scale=0.55]{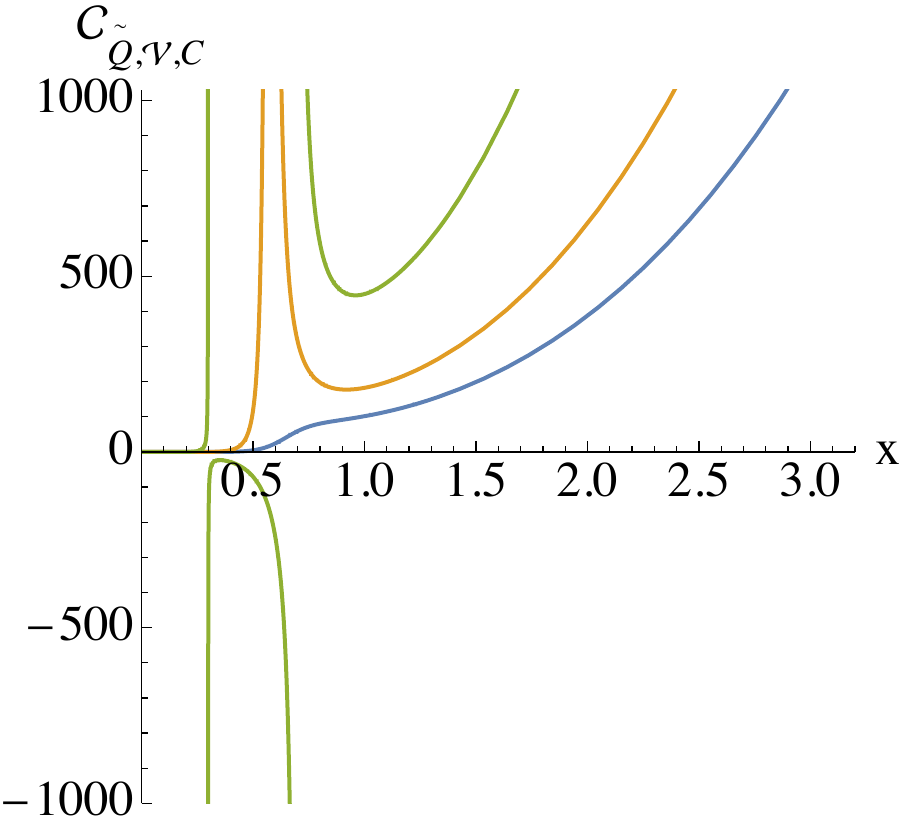}
    \includegraphics[scale=0.55]{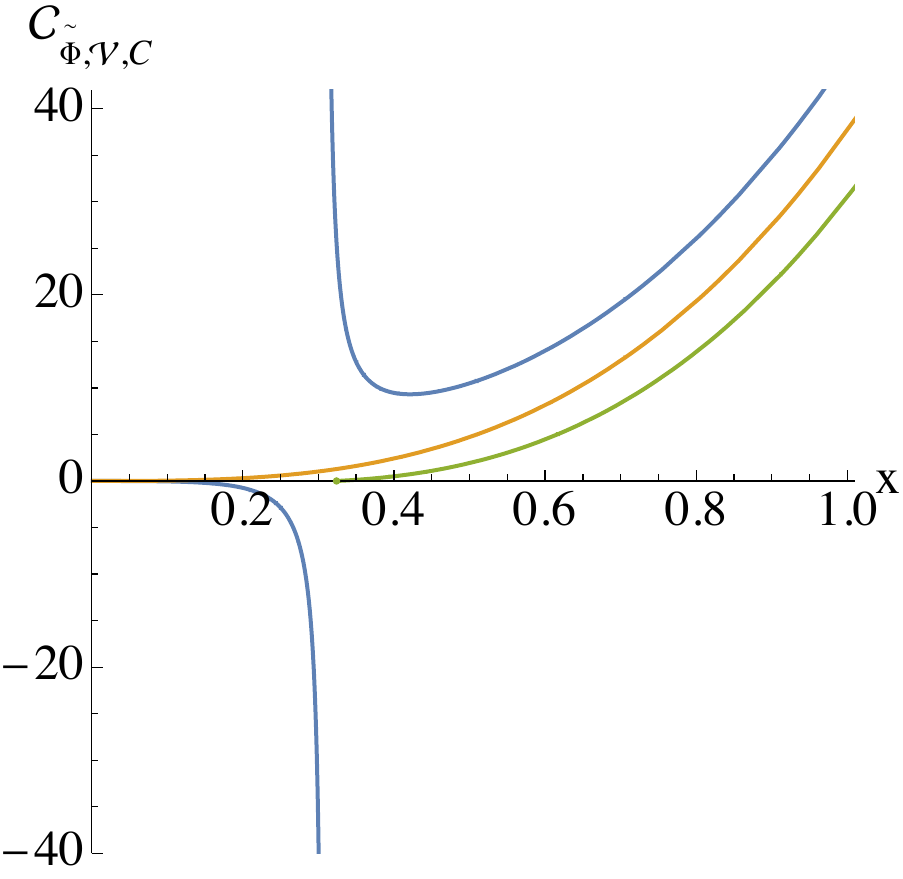}
    \includegraphics[scale=0.55]{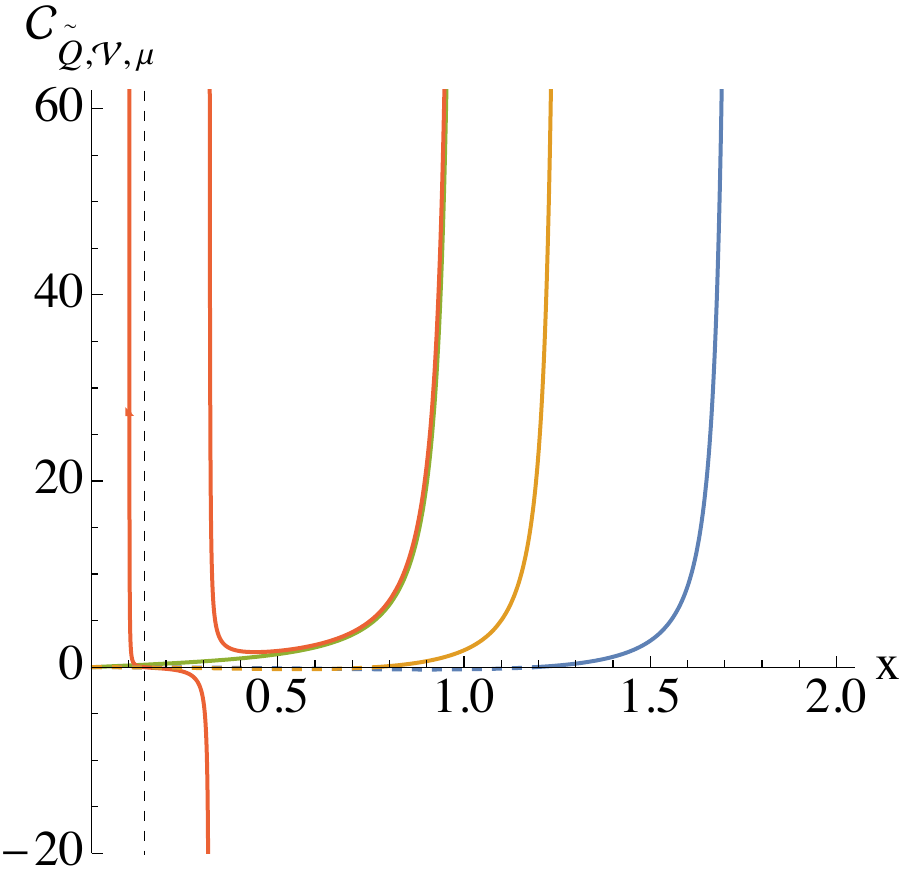}
    \caption{Heat capacities, from left to right,  for fixed $(\tilde Q,{\cal V},C)$, $(\tilde\Phi ,{\cal V},C)$ and $(\tilde Q,{\cal V},\mu)$ ensembles in $d=4$. For ease of  comparison, the parameters used here are the same as  in Figures \ref{fig:qvcFT}, \ref{fig:WTplot} and \ref{fig:GTplot}, respectively.}
    \label{fig:HC}
\end{figure}
We turn next to   the thermodynamic stability of the phases in the three  main ensembles considered in section \ref{sec:ensembles}. The heat capacities in the different ensembles are defined as
 \begin{align}
 	{\cal C}_{\tilde Q,{\cal V},C} &\equiv T \left ( \frac{\partial S}{\partial T}\right)_{\tilde Q, {\cal V}, C} =  \left ( \frac{\partial E }{\partial T }\right)_{\tilde Q, {\cal V}, C}\,,\\
  {\cal C}_{\tilde\Phi,{\cal V},C}  &\equiv T \left ( \frac{\partial S}{\partial T}\right)_{\tilde\Phi, {\cal V}, C} =   \left( \frac{\partial (E- \tilde\Phi \tilde Q) }{ \partial T   }  \right)_{\tilde\Phi, {\cal V}, C}\,,\\
 	{\cal C}_{\tilde Q,{\cal V},\mu}  &\equiv T \left ( \frac{\partial S}{\partial T}\right)_{\tilde Q, {\cal V}, \mu} = \left ( \frac{\partial (E- \mu C) }{ \partial T  }  \right)_{\tilde Q, {\cal V}, \mu}\,.
\end{align} 

We can compute the heat capacities explicitly by first finding the expressions for temperature and entropy as functions of $(\tilde Q, {\cal V},C,x)$, $(\tilde \Phi, {\cal V}, C,x)$  and $(\tilde Q, {\cal V}, \mu, x)$, respectively, and then differentiating using the chain rule. This leads to the following expressions
 \begin{align}
 	{\cal C}_{\tilde Q,{\cal V},C} &= \frac{4 \pi  C (d-1) \left(8 C^2 (d-1) x^{2 d} \left(d x^2+d-2\right)-\tilde Q^2 x^4\right)}{8 C^2 (d-1) \left(d \left(x^2-1\right)+2\right) x^{d+1}+(2
   d-3) \tilde Q^2 x^{-(d-5)}}\,, \\
   	{\cal C}_{\tilde\Phi,{\cal V},C} &= \frac{4 \pi  C (d-1) x^{d-1} \left((d-1) \left(d x^2+d-2\right)-2 (d-2)^2 R^2 \tilde\Phi^2\right)}{d^2 (x^2-1)+2 (d-2)^2 R^2 \tilde\Phi^2-d x^2+3 d-2}\,,\\
  	{\cal C}_{\tilde Q,{\cal V},\mu} &= \frac{\pi  \tilde Q x^{\frac{d}{2}+1} \left(2 x^d-d \mu  R x^2\right) \left((d-2) \mu  R+2 (d-1) x^d\right)}{\sqrt{2 (d-2)} (d-1)^{3/2} \left(2
   x^d-(d-2) \mu  R\right) \left(-x^{d+2}+x^d-\mu  R x^2\right)^{3/2}}\,,
   \end{align}
where we note that $R$ is related to $\cal V$ through equation \eqref{vol}.

 These heat capacities are plotted in Figure \ref{fig:HC} for $d=4$. The left plot corresponds to  the fixed $(\tilde Q,{\cal V},C)$ ensemble, for which the heat capacity behaves as follows. For $C>C_{crit}$ (green), the curve separates into three distinct segments, namely the   low-entropy, intermediate and high-entropy states of the CFT. These three segments correspond to the three branches in the swallowtail $F(T)$ plot in Figure \ref{fig:qvcFT}. The low- and high-entropy CFT states have positive heat capacities and are stable, while the intermediate entropy state has negative heat capacity and is unstable. At $C=C_{crit}$ (orange), the heat capacity is always positive, and becomes infinite at the critical point $x_{crit}=\frac{d-2}{\sqrt{d(d-1)}}$. For $C<C_{crit}$ (blue) the CFT does not display distinct thermodynamic phases and the heat capacity is always positive.
 
 The middle plot of Figure \ref{fig:HC} corresponds to the fixed $(\tilde \Phi, {\cal V},C)$ ensemble. When $\tilde \Phi<\tilde \Phi_c$ (blue), the curve splits into two segments: at small $x$, corresponding to the low-entropy branch  of the blue curve in Figure \ref{fig:WTplot}, the CFT has negative heat capacity and is unstable; for big $x>x_{cusp}=\sqrt{\frac{d^2-2 (d-2)^2 R^2 \tilde \Phi^2-3 d+2}{(d-1) d}}$,  corresponding to the high-entropy branch of the blue curve in Figure \ref{fig:WTplot}, the heat capacity is positive  and the CFT is in a stable state. The heat capacity diverges at $x=x_{cusp}$, which is identical to   the  value of the cusp in equation \eqref{cusp}. For $\tilde \Phi \ge \tilde \Phi_c$ (orange, green) the heat capacity is always positive.
 
 The last plot of Figure \ref{fig:HC} shows the heat capacity for the fixed $(\tilde Q,{\cal V},\mu)$ ensemble. In this ensemble, the heat capacity is always positive for $\mu \leq 0$. Note that for $\mu<0$ (blue, orange), the physical parameter range is $x\geq\left(\frac{  d-2 }{2(d-1)}|\mu| R \right)^{1/d}$, corresponding to the solid part of the $\mu<0$  curves. The dashed horizontal curve has negative temperature and is unphysical. For $\mu>0$ (red), the plot consists of two disconnected segments corresponding to the low-entropy phase (small $x$) and high-entropy phase (large $x$). While the CFT in the high-entropy phase is always stable with positive heat capacity, the heat capacity of the low-entropy phase cuts the $x$-axis at  $x_{int}=\big(\frac{d}{2}\mu R\big)^{1/(d-2)}$  (black dashed line) and becomes negative beyond this point. On the $G-T$ plot of Figure \ref{fig:GTplot}, low-entropy states on the red curve lying to the right of the black dashed line have ${\cal C}_{\tilde Q,{\cal V},\mu}>0$ while those lying between $T_0$ and the dashed line have ${\cal C}_{\tilde Q,{\cal V},\mu}<0$. Note we computed the temperature $T_{int}$ at which the heat capacity of the low-entropy phase switches sign   in   equation \eqref{intermediatetemp}. The separation between the positive and negative heat capacity low-entropy states is, furthermore, shown on the coexistence diagram of Figure \ref{fig:GTplot} by the horizontal dashed line between the red and yellow regions. Finally,  ${\cal C}_{\tilde Q,{\cal V},\mu}$ diverges at the temperatures $T_0$, $T_1$ and $T_2$, corresponding respectively to the left boundary, the vertical dashed  line and the right boundary in the $\mu-T$ diagram of Figure \ref{fig:GTplot}.

 \section{Discussion}
 \label{sec:discussion}

We have studied the thermodynamics of CFTs in thermal states that are dual to charged AdS black holes via the AdS/CFT correspondence.  The following pairs of conjugate thermodynamic variables --  $\{(T,S),(\tilde \Phi,\tilde Q),(p,\mathcal{V}),(\mu, C)\}$ -- are associated with these thermal states, and  each  has a dual expression in terms of   thermodynamic variables of the bulk AdS black hole through the holographic dictionary. We note in particular that  the pair $(\mu,C)$ has been left out in most thermodynamic studies of holographic CFTs up to now. 
To our knowledge, ours is the first exhaustive study of all possible canonical ensembles 
of the CFT under this new addition. We have not performed   independent computations in the CFT to confirm our results, which is an interesting direction for future work.  

Only in the ensemble with fixed  $(\tilde Q, \mathcal{V},C)$ did we find  critical behaviour. Criticality has been found before  \cite{Dolan:2016jjc} in  the $\tilde Q - \tilde \Phi$ plane for $d=4$ holographic CFTs;   the critical exponents
were also shown to agree with mean field theory when the right order parameter, namely $\tilde \Phi$,  was chosen.  Here we have    considered in addition criticality in the $C-\mu$ plane.  Moreover, we discovered a new zeroth-order phase transition in the fixed $(\tilde Q, {\cal V}, \mu)$ ensemble between a
low-entropy (small $S/C$) and a
high-entropy (large $S/C$) phase. We refer to \cite{Altamirano:2013ane} for a different example of a zeroth-order phase transition in extended black hole thermodynamics, and to \cite{Maslov2004ZerothOrderPT} for such a transition in superfluidity and superconductivity.
  The physical viability of  a zeroth-order transition remains an open question.  
  
 The fixed $\mu$ ensemble, for which we found a zeroth-order phase transition, certainly deserves further study. We would like to remind the reader that the chemical potential is equal to the grand canonical free energy $W=E-TS-\tilde \Phi \tilde Q$ divided by the central charge $C$, i.e. $\mu = W/C$. Hence, fixing $\mu$ amounts to fixing the thermal free energy   per degree of freedom, which we note is  a property of the thermal state under consideration. The ensemble then compares different states with different thermal free energy per degree of freedom.    Further, the central charge   of the CFT is allowed to vary continuously in this ensemble, which for gauge theories only makes sense in the large-$N$ limit. It would be interesting to better understand  how the rank  $N$ of the gauge group can be  allowed to vary while keeping $\mu$ fixed in, for instance, $\mathcal N=4$   supersymmetric Yang-Mills theory. However, in the large-$N$ limit the  $N$ dependence of the thermodynamic quantities like energy, entropy and charge is trivial: they just scale with a power of $N$, for instance $N^2$ for $SU(N)$ theories. Thus, changing $N$ just rescales the thermodynamic quantities, and therefore in the large-$N$ limit it seems straightforward to study how various physical quantities change as the central charge varies. Of course, this becomes less trivial when $1/N$ corrections are taken into account.  

The earliest studies of the phase behaviour of  charged AdS black holes \cite{Chamblin:1999tk,Chamblin:1999hg} were performed   well before variations of $\Lambda$ and $G_N$ had been introduced, and so these quantities were implicitly kept constant in the analysis. It is clear that fixing 
$\Lambda$ and $G_N$ corresponds to fixing the    central charge $C$ in the dual CFT; in addition the boundary volume $\cal V$ was   kept fixed in  the holographic interpretation. The canonical (fixed charge) and grand canonical (fixed potential) ensembles studied previously in the bulk   \cite{Chamblin:1999tk,Chamblin:1999hg} are therefore respectively identical to the $(\tilde Q, {\cal V}, C)$ and $(\tilde \Phi, {\cal V},C)$ ensembles in the dual CFT that we have considered.  In particular, the Van der Waals-like phase transition with mean field critical exponents in the bulk  canonical ensemble is the same as the first-order phase transition in the $(\tilde Q, {\cal V}, C)$ ensemble studied here, and the generalised Hawking-Page transition in the grand canonical ensemble in the bulk is dual to the (de)confinement phase transition in the $(\tilde \Phi , {\cal V}, C)$ ensemble. In particular, the free energy plots in Figures $4$ and $3$ in \cite{Chamblin:1999hg}  in the   canonical and grand canonical ensembles  are respectively similar to  our Figures \ref{fig:qvcFT} and~\ref{fig:WTplot} for the $(\tilde Q, {\cal V},C)$ and $(\tilde \Phi,{\cal V},C)$ ensembles. Moreover, the schematic coexistence diagrams in   Figure~$12$ of \cite{Chamblin:1999hg}    qualitatively agree with our exact phase diagrams   in Figures   \ref{fig:WTplot} and \ref{fig:qvcFT2}.

Since the thermodynamics of AdS black holes has been extended to allow for variations of $\Lambda$ (and later on for variations of $G_N$), a natural follow-up question concerns the dual CFT interpretation   of this extended  thermodynamics\footnote{For other field theory interpretations of extended black hole thermodynamics, see e.g. \cite{Johnson:2014yja,Caceres:2015vsa,Dolan:2016jjc,Couch:2016exn,Johnson:2019wcq,Rafiee:2021hyj}. }.  The original investigations \cite{Kastor:2009wy,Dolan:2014cja,Kastor:2014dra,Johnson:2014yja}  already recognized that the cosmological constant is dual to the number of colors or the central charge of the dual CFT. 
However, in previous holographic proposals for   extended black hole thermodynamics   where the central charge and its associated chemical potential were introduced, 
the  CFT volume $\cal V$ and the central charge $C$ were often not clearly distinguished in the bulk, except in   \cite{Karch:2015rpa, Visser:2021eqk}. For instance in \cite{Dolan:2014cja,Zhang:2014uoa,Zhang:2015ova}, the chemical potential was defined as $\mu \equiv\left ( \frac{\partial E}{\partial C} \right)_{S, \tilde Q}$ without fixing the CFT volume $\cal V$. This is problematic since they worked with a particular choice of CFT metric for which both $\cal V$ and $C$ are proportional to $L^{d-1}$ in the bulk, so one has to be careful with taking partial derivatives with respect to $C$ while holding $\cal V$ fixed.  One way to resolve this is to allow for variations of Newton's constant such that the variation of ${\cal V} \sim L^{d-1}$ is   clearly distinct from the variation of $C \sim L^{d-1}/G_N$~\cite{Karch:2015rpa}. Another way to more clearly distinguish $\cal V$ and $C$  is to introduce   the boundary curvature radius $R$, which can be unequal to the bulk curvature radius $L$,  such that the volume is ${\cal V} \sim R^{d-1}$ and the central charge is still $C \sim L^{d-1}/G_N$ \cite{Visser:2021eqk}. Note that in this case $G_N$ does not need to be varied. To cover the most general case in the present paper, following \cite{Visser:2021eqk}, we allowed for  variations of Newton's constant and considered a constant boundary curvature radius $R \neq L.$ 

 Recently a   ``mixed'' bulk/boundary perspective on extended black hole thermodynamics was considered by some of us \cite{Cong:2021fnf}, where    the variation of both the  bulk pressure and of the central charge   appear in the extended first law of   AdS black holes (see footnote \ref{footnote1}). 
The fixed $(Q_b,P,C)$ ensemble for charged AdS black holes considered in that paper corresponds to the fixed $(\tilde Q,{\cal V}, C)$ ensemble   in the present paper. Similar critical behaviour was discovered for $d=3$ in the $C-\mu$ plane in  that paper. One important difference, however, is that the coexistence line in the $1/C-T$ phase diagram in \cite{Cong:2021fnf} has positive slope and agrees with the coexistence line in the $P-T$ phase diagram of the Van der Waals liquid-gas system, whereas the coexistence curve in the $1/C-T$ phase diagram   in Figure \ref{fig:qvcFT2} has negative slope and does not start from the origin.  Moreover, another difference is that at the critical point the central charge is proportional to the square of the   electric charge  $C_{crit}\sim Q_b^2$ in $d=3$, cf. equation (28) in~\cite{Cong:2021fnf}, which is seemingly at odds with our result \eqref{Ccrit}, $C_{crit}\sim \tilde Q$. This is not a disagreement,   because the $Q_b$ appearing in that formula is the \textit{bulk} electric charge,  and
the holographic dictionary employed in \cite{Cong:2021fnf}, $\tilde Q   = Q_b L / \sqrt{G_N}$,  alters the relationship    between
the electric charge and central charge.  Indeed, if we make use of this latter  dictionary definition, and note that $C \sim L^2/G_N$ for $d=3$, then upon insertion into $C_{crit} \sim Q_b^2$   we would    obtain $C_{crit}\sim \tilde Q $ in agreement with our result\footnote{In general dimensions the critical central charge found in \cite{Cong:2021fnf} is given by   $C_{crit} \sim G_N^{- \frac{d-3}{2d-4}} Q_b^{\frac{d-1}{d-2}}$. Inserting the dictionary    $\tilde Q = Q_b  L / \sqrt{G_N}$ and $C\sim L^{d-1}/G_N$  still   yields $C_{crit} \sim  \tilde Q,$ in agreement with our equation \eqref{Ccrit}. }.

In other recent work \cite{Zeyuan:2021uol} a restricted version of the   CFT thermodynamics in \cite{Visser:2021eqk} was proposed, where the volume $\cal V$ was kept fixed.  The reason for this seems to be that the authors contend the boundary volume is related to the cosmological constant in the bulk, via ${\cal V}\sim L^{d-1}$,  which in turn is kept fixed in order to avoid changing the gravitational theory. However in section \ref{sec:holographic} (and in \cite{Visser:2021eqk}), we explained that the holographic dictionary for the volume can be altered into   ${\cal V}\sim R^{d-1}$  by Weyl rescaling the CFT metric, where $R$ is not related to the cosmological constant.  Furthermore, the  claim in \cite{Zeyuan:2021uol}  that the absence of  a $p {\cal V}$  term in the holographic Euler equation is a problem for the homogeneity behaviour of the internal energy is not correct. The Euler equation is derived from the scaling relation   $E(\alpha S, V, \alpha \tilde Q, \alpha C) = \alpha E(S,V, \tilde Q, C)$, which holds in the deconfined phase of a large-$N$ theory on a compact space. This means the internal energy is not extensive in this setting, since it does not satisfy $E(\alpha S, \alpha V, \alpha \tilde Q ,C) = \alpha E(S,V, \tilde Q,C )$. Only in the limit $TR \to \infty$ does the energy become an extensive function.

Finally, we   comment on the different findings in the literature  regarding criticality for charged AdS black holes. In standard  black hole chemistry, where Newton's constant is kept fixed,  criticality is present    in the $Q-\Phi$ plane, the $\beta - r_h$ plane and   the $P-V$ plane for charged AdS black holes \cite{Chamblin:1999tk,Chamblin:1999hg,Kubiznak:2012wp}. The pleasant feature of $P-V$ criticality  is that the analogy with the Van der Waals fluid is complete, since it compares the ``right'' physical quantities in the black hole system and the liquid-gas system. 
However, as explained in the introduction, when Newton's constant is allowed to vary, the extended first law cannot be expressed solely in terms of the variation of the bulk pressure. Therefore, in \cite{Cong:2021fnf} the extended  black hole first law was written in a ``mixed'' way in terms of  the bulk pressure and central charge variations. In this ``mixed'' bulk/boundary formalism there is no criticality in the $P-V$ plane, but a new kind of criticality was discovered in the $C-\mu$ plane. Furthermore, in the present paper  we studied a CFT interpretation of extended black hole thermodynamics and found criticality in the $\tilde Q - \tilde \Phi$ plane and in the $C-\mu$ plane (where our chemical potential is notably different from the one in \cite{Cong:2021fnf}). Thus, on the one hand we see that   $P-V$ criticality is only present in the bulk when Newton's constant is kept fixed. On the other hand, $P-V$ criticality disappears   when Newton's constant is allowed to vary, and it is traded for $C-\mu$ criticality. Similarly, our findings in the CFT     suggest    there is no $p-\cal V$ criticality, but there is a critical point in the $C-\mu$ plane that falls within the same universality class as the Van der
Waals fluid.

\section*{Acknowledgments}
\noindent MV is funded by the Swiss National Science Foundation, through Project Grants 200020-182513 and 51NF40-141869 The Mathematics of Physics (SwissMAP). This work was supported by the Perimeter Institute for Theoretical Physics and by NSERC. Research at Perimeter Institute is supported in part by the Government of Canada through the Department of Innovation, Science and Economic Development Canada and by the Province of Ontario through the Ministry of Colleges and Universities. Perimeter Institute and the University of Waterloo are situated on the Haldimand Tract, land that was promised to the Haudenosaunee of the Six Nations of the Grand River, and is within the territory of the Neutral, Anishnawbe, and Haudenosaunee peoples.

\bibliography{criticality}

\providecommand{\href}[2]{#2}\begingroup\raggedright\begin{thebibliography}{10}

\bibitem{hawking1975particle}
S.~W. Hawking, {\it Particle creation by black holes},  {\em Communications in
  mathematical physics} {\bf 43} (1975), no.~3 199--220.

\bibitem{Bekenstein:1973ur}
J.~D. Bekenstein, {\it {Black holes and entropy}},  {\em Phys. Rev. D} {\bf 7}
  (1973) 2333--2346.

\bibitem{Maldacena:1997re}
J.~M. Maldacena, {\it {The Large N limit of superconformal field theories and
  supergravity}},  {\em Adv. Theor. Math. Phys.} {\bf 2} (1998) 231--252,
  [\href{http://xxx.arxiv.org/abs/hep-th/9711200}{{\tt hep-th/9711200}}].

\bibitem{Hawking:1982dh}
S.~W. Hawking and D.~N. Page, {\it {Thermodynamics of Black Holes in anti-De
  Sitter Space}},  {\em Commun. Math. Phys.} {\bf 87} (1983) 577.

\bibitem{Witten:1998zw}
E.~Witten, {\it {Anti-de Sitter space, thermal phase transition, and
  confinement in gauge theories}},  {\em Adv. Theor. Math. Phys.} {\bf 2}
  (1998) 505--532, [\href{http://xxx.arxiv.org/abs/hep-th/9803131}{{\tt
  hep-th/9803131}}].

\bibitem{Chamblin:1999tk}
A.~Chamblin, R.~Emparan, C.~V. Johnson, and R.~C. Myers, {\it {Charged AdS
  black holes and catastrophic holography}},  {\em Phys. Rev. D} {\bf 60}
  (1999) 064018, [\href{http://xxx.arxiv.org/abs/hep-th/9902170}{{\tt
  hep-th/9902170}}].

\bibitem{Chamblin:1999hg}
A.~Chamblin, R.~Emparan, C.~V. Johnson, and R.~C. Myers, {\it {Holography,
  thermodynamics and fluctuations of charged AdS black holes}},  {\em Phys.
  Rev. D} {\bf 60} (1999) 104026,
  [\href{http://xxx.arxiv.org/abs/hep-th/9904197}{{\tt hep-th/9904197}}].

\bibitem{Cvetic:1999ne}
M.~Cvetic and S.~S. Gubser, {\it {Phases of R charged black holes, spinning
  branes and strongly coupled gauge theories}},  {\em JHEP} {\bf 04} (1999)
  024, [\href{http://xxx.arxiv.org/abs/hep-th/9902195}{{\tt hep-th/9902195}}].

\bibitem{Kubiznak:2012wp}
D.~Kubiznak and R.~B. Mann, {\it {P-V criticality of charged AdS black holes}},
   {\em JHEP} {\bf 07} (2012) 033,
  [\href{http://xxx.arxiv.org/abs/1205.0559}{{\tt arXiv:1205.0559}}].

\bibitem{Dolan:2014vba}
B.~P. Dolan, A.~Kostouki, D.~Kubiznak, and R.~B. Mann, {\it {Isolated critical
  point from Lovelock gravity}},  {\em Class. Quant. Grav.} {\bf 31} (2014),
  no.~24 242001, [\href{http://xxx.arxiv.org/abs/1407.4783}{{\tt
  arXiv:1407.4783}}].

\bibitem{Altamirano:2013ane}
N.~Altamirano, D.~Kubiznak, and R.~B. Mann, {\it {Reentrant phase transitions
  in rotating anti\textendash{}de Sitter black holes}},  {\em Phys. Rev. D}
  {\bf 88} (2013), no.~10 101502,
  [\href{http://xxx.arxiv.org/abs/1306.5756}{{\tt arXiv:1306.5756}}].

\bibitem{Frassino:2014pha}
A.~M. Frassino, D.~Kubiznak, R.~B. Mann, and F.~Simovic, {\it {Multiple
  Reentrant Phase Transitions and Triple Points in Lovelock Thermodynamics}},
  {\em JHEP} {\bf 09} (2014) 080,
  [\href{http://xxx.arxiv.org/abs/1406.7015}{{\tt arXiv:1406.7015}}].

\bibitem{Altamirano:2013uqa}
N.~Altamirano, D.~Kubiz\v{n}\'ak, R.~B. Mann, and Z.~Sherkatghanad, {\it
  {Kerr-AdS analogue of triple point and solid/liquid/gas phase transition}},
  {\em Class. Quant. Grav.} {\bf 31} (2014) 042001,
  [\href{http://xxx.arxiv.org/abs/1308.2672}{{\tt arXiv:1308.2672}}].

\bibitem{Wei:2014hba}
S.-W. Wei and Y.-X. Liu, {\it {Triple points and phase diagrams in the extended
  phase space of charged Gauss-Bonnet black holes in AdS space}},  {\em Phys.
  Rev. D} {\bf 90} (2014), no.~4 044057,
  [\href{http://xxx.arxiv.org/abs/1402.2837}{{\tt arXiv:1402.2837}}].

\bibitem{Hennigar:2016xwd}
R.~A. Hennigar, R.~B. Mann, and E.~Tjoa, {\it {Superfluid Black Holes}},  {\em
  Phys. Rev. Lett.} {\bf 118} (2017), no.~2 021301,
  [\href{http://xxx.arxiv.org/abs/1609.02564}{{\tt arXiv:1609.02564}}].

\bibitem{Kastor:2009wy}
D.~Kastor, S.~Ray, and J.~Traschen, {\it {Enthalpy and the Mechanics of AdS
  Black Holes}},  {\em Class. Quant. Grav.} {\bf 26} (2009) 195011,
  [\href{http://xxx.arxiv.org/abs/0904.2765}{{\tt arXiv:0904.2765}}].

\bibitem{Kastor:2010gq}
D.~Kastor, S.~Ray, and J.~Traschen, {\it {Smarr Formula and an Extended First
  Law for Lovelock Gravity}},  {\em Class. Quant. Grav.} {\bf 27} (2010)
  235014, [\href{http://xxx.arxiv.org/abs/1005.5053}{{\tt arXiv:1005.5053}}].

\bibitem{Kastor:2014dra}
D.~Kastor, S.~Ray, and J.~Traschen, {\it {Chemical Potential in the First Law
  for Holographic Entanglement Entropy}},  {\em JHEP} {\bf 11} (2014) 120,
  [\href{http://xxx.arxiv.org/abs/1409.3521}{{\tt arXiv:1409.3521}}].

\bibitem{Karch:2015rpa}
A.~Karch and B.~Robinson, {\it {Holographic Black Hole Chemistry}},  {\em JHEP}
  {\bf 12} (2015) 073, [\href{http://xxx.arxiv.org/abs/1510.02472}{{\tt
  arXiv:1510.02472}}].

\bibitem{Sarkar:2020yjs}
D.~Sarkar and M.~Visser, {\it {The first law of differential entropy and
  holographic complexity}},  {\em JHEP} {\bf 11} (2020) 004,
  [\href{http://xxx.arxiv.org/abs/2008.12673}{{\tt arXiv:2008.12673}}].

\bibitem{Visser:2021eqk}
M.~R. Visser, {\it {Holographic Thermodynamics Requires a Chemical Potential
  for Color}},  \href{http://xxx.arxiv.org/abs/2101.04145}{{\tt
  arXiv:2101.04145}}.

\bibitem{Cong:2021fnf}
W.~Cong, D.~Kubiznak, and R.~B. Mann, {\it {Thermodynamics of AdS Black Holes:
  Critical Behavior of the Central Charge}},  {\em Phys. Rev. Lett.} {\bf 127}
  (2021), no.~9 091301, [\href{http://xxx.arxiv.org/abs/2105.02223}{{\tt
  arXiv:2105.02223}}].

\bibitem{Jacobson:2018ahi}
T.~Jacobson and M.~Visser, {\it {Gravitational Thermodynamics of Causal
  Diamonds in (A)dS}},  {\em SciPost Phys.} {\bf 7} (2019), no.~6 079,
  [\href{http://xxx.arxiv.org/abs/1812.01596}{{\tt arXiv:1812.01596}}].

\bibitem{Dolan:2011xt}
B.~P. Dolan, {\it {Pressure and volume in the first law of black hole
  thermodynamics}},  {\em Class. Quant. Grav.} {\bf 28} (2011) 235017,
  [\href{http://xxx.arxiv.org/abs/1106.6260}{{\tt arXiv:1106.6260}}].

\bibitem{Dolan:2010ha}
B.~P. Dolan, {\it {The cosmological constant and the black hole equation of
  state}},  {\em Class. Quant. Grav.} {\bf 28} (2011) 125020,
  [\href{http://xxx.arxiv.org/abs/1008.5023}{{\tt arXiv:1008.5023}}].

\bibitem{Cvetic:2010jb}
M.~Cvetic, G.~W. Gibbons, D.~Kubiznak, and C.~N. Pope, {\it {Black Hole
  Enthalpy and an Entropy Inequality for the Thermodynamic Volume}},  {\em
  Phys. Rev. D} {\bf 84} (2011) 024037,
  [\href{http://xxx.arxiv.org/abs/1012.2888}{{\tt arXiv:1012.2888}}].

\bibitem{Kubiznak:2014zwa}
D.~Kubiznak and R.~B. Mann, {\it {Black hole chemistry}},  {\em Can. J. Phys.}
  {\bf 93} (2015), no.~9 999--1002,
  [\href{http://xxx.arxiv.org/abs/1404.2126}{{\tt arXiv:1404.2126}}].

\bibitem{Johnson:2014yja}
C.~V. Johnson, {\it {Holographic Heat Engines}},  {\em Class. Quant. Grav.}
  {\bf 31} (2014) 205002, [\href{http://xxx.arxiv.org/abs/1404.5982}{{\tt
  arXiv:1404.5982}}].

\bibitem{Dolan:2014cja}
B.~P. Dolan, {\it {Bose condensation and branes}},  {\em JHEP} {\bf 10} (2014)
  179, [\href{http://xxx.arxiv.org/abs/1406.7267}{{\tt arXiv:1406.7267}}].

\bibitem{Zhang:2014uoa}
J.-L. Zhang, R.-G. Cai, and H.~Yu, {\it {Phase transition and thermodynamical
  geometry for Schwarzschild AdS black hole in AdS$_{5}$ \texttimes{} S$^{5}$
  spacetime}},  {\em JHEP} {\bf 02} (2015) 143,
  [\href{http://xxx.arxiv.org/abs/1409.5305}{{\tt arXiv:1409.5305}}].

\bibitem{Zhang:2015ova}
J.-L. Zhang, R.-G. Cai, and H.~Yu, {\it {Phase transition and thermodynamical
  geometry of Reissner-Nordstr\"om-AdS black holes in extended phase space}},
  {\em Phys. Rev. D} {\bf 91} (2015), no.~4 044028,
  [\href{http://xxx.arxiv.org/abs/1502.01428}{{\tt arXiv:1502.01428}}].

\bibitem{Dolan:2016jjc}
B.~P. Dolan, {\it {Pressure and compressibility of conformal field theories
  from the AdS/CFT correspondence}},  {\em Entropy} {\bf 18} (2016) 169,
  [\href{http://xxx.arxiv.org/abs/1603.06279}{{\tt arXiv:1603.06279}}].

\bibitem{Pedraza:2021cvx}
J.~F. Pedraza, A.~Svesko, W.~Sybesma, and M.~R. Visser, {\it {Semi-classical
  thermodynamics of quantum extremal surfaces in Jackiw-Teitelboim gravity}},
  \href{http://xxx.arxiv.org/abs/2107.10358}{{\tt arXiv:2107.10358}}.

\bibitem{Gauntlett:1998fz}
J.~P. Gauntlett, R.~C. Myers, and P.~K. Townsend, {\it {Black holes of D = 5
  supergravity}},  {\em Class. Quant. Grav.} {\bf 16} (1999) 1--21,
  [\href{http://xxx.arxiv.org/abs/hep-th/9810204}{{\tt hep-th/9810204}}].

\bibitem{Caldarelli:1999xj}
M.~M. Caldarelli, G.~Cognola, and D.~Klemm, {\it {Thermodynamics of
  Kerr-Newman-AdS black holes and conformal field theories}},  {\em Class.
  Quant. Grav.} {\bf 17} (2000) 399--420,
  [\href{http://xxx.arxiv.org/abs/hep-th/9908022}{{\tt hep-th/9908022}}].

\bibitem{Kubiznak:2016qmn}
D.~Kubiznak, R.~B. Mann, and M.~Teo, {\it {Black hole chemistry: thermodynamics
  with Lambda}},  {\em Class. Quant. Grav.} {\bf 34} (2017), no.~6 063001,
  [\href{http://xxx.arxiv.org/abs/1608.06147}{{\tt arXiv:1608.06147}}].

\bibitem{Caceres:2016xjz}
E.~Caceres, P.~H. Nguyen, and J.~F. Pedraza, {\it {Holographic entanglement
  chemistry}},  {\em Phys. Rev. D} {\bf 95} (2017), no.~10 106015,
  [\href{http://xxx.arxiv.org/abs/1605.00595}{{\tt arXiv:1605.00595}}].

\bibitem{Rosso:2020zkk}
F.~Rosso and A.~Svesko, {\it {Novel aspects of the extended first law of
  entanglement}},  {\em JHEP} {\bf 08} (2020), no.~08 008,
  [\href{http://xxx.arxiv.org/abs/2003.10462}{{\tt arXiv:2003.10462}}].

\bibitem{Sinamuli:2017rhp}
M.~Sinamuli and R.~B. Mann, {\it {Higher Order Corrections to Holographic Black
  Hole Chemistry}},  {\em Phys. Rev. D} {\bf 96} (2017), no.~8 086008,
  [\href{http://xxx.arxiv.org/abs/1706.04259}{{\tt arXiv:1706.04259}}].

\bibitem{Brown:1986nw}
J.~Brown and M.~Henneaux, {\it {Central Charges in the Canonical Realization of
  Asymptotic Symmetries: An Example from Three-Dimensional Gravity}},  {\em
  Commun. Math. Phys.} {\bf 104} (1986) 207--226.

\bibitem{Myers:2010xs}
R.~C. Myers and A.~Sinha, {\it {Seeing a c-theorem with holography}},  {\em
  Phys. Rev. D} {\bf 82} (2010) 046006,
  [\href{http://xxx.arxiv.org/abs/1006.1263}{{\tt arXiv:1006.1263}}].

\bibitem{Myers:2010tj}
R.~C. Myers and A.~Sinha, {\it {Holographic c-theorems in arbitrary
  dimensions}},  {\em JHEP} {\bf 01} (2011) 125,
  [\href{http://xxx.arxiv.org/abs/1011.5819}{{\tt arXiv:1011.5819}}].

\bibitem{Osborn:1993cr}
H.~Osborn and A.~C. Petkou, {\it {Implications of conformal invariance in field
  theories for general dimensions}},  {\em Annals Phys.} {\bf 231} (1994)
  311--362, [\href{http://xxx.arxiv.org/abs/hep-th/9307010}{{\tt
  hep-th/9307010}}].

\bibitem{Hung:2011nu}
L.-Y. Hung, R.~C. Myers, M.~Smolkin, and A.~Yale, {\it {Holographic
  Calculations of Renyi Entropy}},  {\em JHEP} {\bf 12} (2011) 047,
  [\href{http://xxx.arxiv.org/abs/1110.1084}{{\tt arXiv:1110.1084}}].

\bibitem{Gubser:1998bc}
S.~S. Gubser, I.~R. Klebanov, and A.~M. Polyakov, {\it {Gauge theory
  correlators from noncritical string theory}},  {\em Phys. Lett. B} {\bf 428}
  (1998) 105--114, [\href{http://xxx.arxiv.org/abs/hep-th/9802109}{{\tt
  hep-th/9802109}}].

\bibitem{Witten:1998qj}
E.~Witten, {\it {Anti-de Sitter space and holography}},  {\em Adv. Theor. Math.
  Phys.} {\bf 2} (1998) 253--291,
  [\href{http://xxx.arxiv.org/abs/hep-th/9802150}{{\tt hep-th/9802150}}].

\bibitem{Savonije:2001nd}
I.~Savonije and E.~P. Verlinde, {\it {CFT and entropy on the brane}},  {\em
  Phys. Lett. B} {\bf 507} (2001) 305--311,
  [\href{http://xxx.arxiv.org/abs/hep-th/0102042}{{\tt hep-th/0102042}}].

\bibitem{Henningson:1998gx}
M.~Henningson and K.~Skenderis, {\it {The Holographic Weyl anomaly}},  {\em
  JHEP} {\bf 07} (1998) 023,
  [\href{http://xxx.arxiv.org/abs/hep-th/9806087}{{\tt hep-th/9806087}}].

\bibitem{Balasubramanian:1999re}
V.~Balasubramanian and P.~Kraus, {\it {A Stress tensor for Anti-de Sitter
  gravity}},  {\em Commun. Math. Phys.} {\bf 208} (1999) 413--428,
  [\href{http://xxx.arxiv.org/abs/hep-th/9902121}{{\tt hep-th/9902121}}].

\bibitem{Henneaux:1985tv}
M.~Henneaux and C.~Teitelboim, {\it {Asymptotically anti-De Sitter Spaces}},
  {\em Commun. Math. Phys.} {\bf 98} (1985) 391--424.

\bibitem{Ashtekar:1984zz}
A.~Ashtekar and A.~Magnon, {\it {Asymptotically anti-de Sitter space-times}},
  {\em Class. Quant. Grav.} {\bf 1} (1984) L39--L44.

\bibitem{Ashtekar:1999jx}
A.~Ashtekar and S.~Das, {\it {Asymptotically Anti-de Sitter space-times:
  Conserved quantities}},  {\em Class. Quant. Grav.} {\bf 17} (2000) L17--L30,
  [\href{http://xxx.arxiv.org/abs/hep-th/9911230}{{\tt hep-th/9911230}}].

\bibitem{Hollands:2005wt}
S.~Hollands, A.~Ishibashi, and D.~Marolf, {\it {Comparison between various
  notions of conserved charges in asymptotically AdS-spacetimes}},  {\em Class.
  Quant. Grav.} {\bf 22} (2005) 2881--2920,
  [\href{http://xxx.arxiv.org/abs/hep-th/0503045}{{\tt hep-th/0503045}}].

\bibitem{Tarrio:2011de}
J.~Tarrio and S.~Vandoren, {\it {Black holes and black branes in Lifshitz
  spacetimes}},  {\em JHEP} {\bf 09} (2011) 017,
  [\href{http://xxx.arxiv.org/abs/1105.6335}{{\tt arXiv:1105.6335}}].

\bibitem{Pedraza:2018eey}
J.~F. Pedraza, W.~Sybesma, and M.~R. Visser, {\it {Hyperscaling violating black
  holes with spherical and hyperbolic horizons}},  {\em Class. Quant. Grav.}
  {\bf 36} (2019), no.~5 054002,
  [\href{http://xxx.arxiv.org/abs/1807.09770}{{\tt arXiv:1807.09770}}].

\bibitem{Niu:2011tb}
C.~Niu, Y.~Tian, and X.-N. Wu, {\it {Critical Phenomena and Thermodynamic
  Geometry of RN-AdS Black Holes}},  {\em Phys. Rev. D} {\bf 85} (2012) 024017,
  [\href{http://xxx.arxiv.org/abs/1104.3066}{{\tt arXiv:1104.3066}}].

\bibitem{Maslov2004ZerothOrderPT}
V.~P. Maslov, {\it Zeroth-order phase transitions},  {\em Mathematical Notes}
  {\bf 76} (2004) 697--710.

\bibitem{Caceres:2015vsa}
E.~Caceres, P.~H. Nguyen, and J.~F. Pedraza, {\it {Holographic entanglement
  entropy and the extended phase structure of STU black holes}},  {\em JHEP}
  {\bf 09} (2015) 184, [\href{http://xxx.arxiv.org/abs/1507.06069}{{\tt
  arXiv:1507.06069}}].

\bibitem{Couch:2016exn}
J.~Couch, W.~Fischler, and P.~H. Nguyen, {\it {Noether charge, black hole
  volume, and complexity}},  {\em JHEP} {\bf 03} (2017) 119,
  [\href{http://xxx.arxiv.org/abs/1610.02038}{{\tt arXiv:1610.02038}}].

\bibitem{Johnson:2019wcq}
C.~V. Johnson, V.~L. Martin, and A.~Svesko, {\it {Microscopic description of
  thermodynamic volume in extended black hole thermodynamics}},  {\em Phys.
  Rev. D} {\bf 101} (2020), no.~8 086006,
  [\href{http://xxx.arxiv.org/abs/1911.05286}{{\tt arXiv:1911.05286}}].

\bibitem{Rafiee:2021hyj}
M.~Rafiee, S.~A.~H. Mansoori, S.-W. Wei, and R.~B. Mann, {\it {Universal
  criticality of thermodynamic geometry for boundary conformal field theories
  in gauge/gravity duality}},  \href{http://xxx.arxiv.org/abs/2107.08883}{{\tt
  arXiv:2107.08883}}.

\bibitem{Zeyuan:2021uol}
G.~Zeyuan and L.~Zhao, {\it {Restricted phase space thermodynamics for AdS
  black holes via holography}},
  \href{http://xxx.arxiv.org/abs/2112.02386}{{\tt arXiv:2112.02386}}.

\end{thebibliography}\endgroup

\end{document}